\documentclass[aps,pra,twocolumn,superscriptaddress,nofootinbib]{revtex4-2}

\usepackage{amssymb,amsmath,amsthm,color,graphicx,times,graphicx}
\usepackage{hyperref}
\usepackage{subfigure}
\usepackage{times}
\usepackage{graphicx}
\usepackage{overpic}
\usepackage{bbold}
\usepackage{color}
\usepackage{amsmath}
\usepackage{booktabs}
\usepackage{hyperref}
\usepackage{amsmath}
\usepackage{orcidlink}
\hypersetup{
    colorlinks=true,
    linkcolor=black,    
    citecolor=black,    
    urlcolor=blue       
}
\makeatletter
\@ifundefined{pdfsuppresswarningpagegroup}{}{\pdfsuppresswarningpagegroup=1}
\makeatother
\newcommand{\bra}[1]{\langle{#1}|}
\newcommand{\ket}[1]{|{#1}\rangle}

\providecommand{\openone}{\leavevmode\hbox{\small1\kern-4.3pt\normalsize1}}

\theoremstyle{plain}

\theoremstyle{definition}

\begin{document}
\title{Influence of quantum decoherence on the survival of quantumness in neutrino oscillations}

\author{Jilali Loulijat}
\affiliation{LPMS, Faculty of Sciences, University Ibn Tofail, Kenitra, Morocco.}\author{Abdallah Slaoui \orcidlink{0000-0002-5284-3240}}\email{abdallah.slaoui@um5s.net.ma}\affiliation{LPHE-Modeling and Simulation, Faculty of Sciences, Mohammed V University in Rabat, Rabat, Morocco.}\affiliation{Centre of Physics and Mathematics, CPM, Faculty of Sciences, Mohammed V University in Rabat, Rabat, Morocco.}\author{Mohamed Gouighri \orcidlink{0000-0002-9551-0251}}\affiliation{LPMS, Faculty of Sciences, University Ibn Tofail, Kenitra, Morocco.}\author{Berihu Teklu \orcidlink{0000-0001-9280-533X}}\email{berihu.gebrehiwot@ku.ac.ae}\affiliation{College of Computing and Mathematical Sciences, Department of Applied Mathematics and Sciences and Center for Cyber-Physical Systems (C2PS), Khalifa University of Science and Technology, 127788, Abu Dhabi, United Arab Emirates.}

\begin{abstract}
This study examines the dynamics of quantumness in two-flavor neutrino oscillations in the presence of a dephasing channel, using representative oscillation parameters from the KamLAND, MINOS, and Daya Bay experiments. We analyze three complementary quantum-correlation measures — entanglement of formation (EOF), quantum discord (QD), and local quantum uncertainty (LQU) — within an effective two-qubit description. In the unitary case, all three measures display oscillatory behavior controlled by flavor mixing, and the relevant mixing angle strongly shapes their amplitudes. MINOS exhibits the largest correlations because $\theta_{23}$ is close to maximal, KamLAND shows intermediate values associated with the solar sector, and Daya Bay yields smaller correlations due to the relatively small value of $\theta_{13}$. Under dephasing, the off-diagonal coherence terms are suppressed, and the three quantifiers decrease accordingly, while QD remains non-zero in regimes where entanglement is weak. For pure states, LQU satisfies $\mathcal{U}=\mathcal{C}^2$ and therefore tracks entanglement monotonically, whereas QD provides a broader witness of non-classical correlations. These results provide a compact quantum information description of two-flavor neutrino oscillations in both the coherent and dephased regimes. We also quantify the sensitivity of these observables to oscillation and decoherence parameters, showing that their main added value relative to flavor probabilities is their direct response to off-diagonal coherence loss.
\end{abstract}

\maketitle

\section{Introduction}

The discovery of neutrino oscillations has been a major milestone in particle physics, providing direct evidence for physics beyond the original formulation of the Standard Model. First proposed in the 1960s by Pontecorvo and by Maki and collaborators~\cite{Pontecorvo1957,Pontecorvo1958,Maki1962}, neutrino oscillations arise because the three flavor eigenstates---electron ($\nu_e$), muon ($\nu_\mu$), and tau ($\nu_\tau$)---are coherent superpositions of the mass eigenstates $\nu_1$, $\nu_2$, and $\nu_3$. The mismatch between flavor and mass bases is encoded in the unitary Pontecorvo–Maki–Nakagawa–Sakata (PMNS) matrix, parametrized by three mixing angles ($\theta_{12},\theta_{13},\theta_{23}$) and a CP-violating phase $\delta_{\mathrm{CP}}$~\cite{Bilenky2010,Giunti2007,Tanabashi2018}. As a consequence, the probability that a neutrino produced in a given flavor state is detected in the same or a different flavor exhibits characteristic distance- and energy-dependent oscillations. This phenomenon has been robustly established in solar~\cite{Cleveland1998,Ahmad2002}, atmospheric~\cite{Fukuda1998}, reactor~\cite{Eguchi2003}, and accelerator~\cite{Adamson2008,Abe2011} experiments, enabling precise measurements of the mixing angles and mass-squared differences that govern neutrino propagation.

In parallel, developments in quantum information theory have clarified the central role of quantum correlations in physical systems. Quantum entanglement, highlighted initially by the Einstein–Podolsky–Rosen paradox and later quantified through Bell’s inequalities~\cite{Einstein1935,Bell1964}, is now understood as a key resource for quantum communication, computation, and metrology. However, entanglement is not the only signature of nonclassicality: separable mixed states can still display quantum correlations that have no classical counterpart. This has motivated the introduction of more general quantifiers, such as quantum discord (QD)~\cite{Ollivier2001,Henderson2001,Giorda2010,Modi2012} and local quantum uncertainty (LQU)~\cite{Girolami2013}, as well as a broad family of entanglement measures, among which the entanglement of formation (EOF)~\cite{Bennett1996,Wootters,Plenio2007} plays a central role. These tools have proven particularly useful for characterizing noisy, mixed-state scenarios typical of realistic experiments.

Neutrino oscillations provide a natural arena for studying such quantum correlations over macroscopic distances and relativistic energy scales. An initial flavor state is a coherent superposition of mass eigenstates whose time evolution can be mapped onto an effective bipartite (two-qubit) system~\cite{Blasone2009,Banerjee2015}. In this effective description, the flavor–mass mismatch generates quantum correlations between different “modes” during propagation, in close analogy with the entanglement shared between spatially separated subsystems. This framework has been used to investigate entanglement, discord, and related quantities in oscillating neutrino systems, as well as their sensitivity to decoherence, matter effects, and other dynamical features~\cite{Modi2012,Blasone2011,blennow2010entanglement,Alok2016,Dixit2019,Blasone2008,Wang2016}. It thereby offers novel insights into neutrino decoherence, environmental noise, and the fundamental “quantumness” of relativistic fields.

Neutrino quantum decoherence has also been extensively studied in the open-quantum-system/Lindblad framework. Theoretical mechanisms include stochastic matter or gravitational backgrounds, fluctuating media, randomly varying fields, particle backgrounds, radiative-decay-induced decoherence, and quantum-spacetime-noise-induced decoherence~\cite{Loreti1994,Burgess1997,Benatti2005,Dvornikov2021,NievesSahu2019,NievesSahu2020,Stankevich2020,Lichkunov2025,Stankevich2025,Nandi2026SpacetimeNoise}. In particular, the recent open-system treatment of stochastic Planck-scale spacetime fluctuations provides a quantum-gravity-motivated Lindblad description with direct application to neutrino oscillations~\cite{Nandi2026SpacetimeNoise}. Phenomenological and experimental analyses have constrained dissipative parameters using solar, reactor, atmospheric, accelerator, IceCube/KM3NeT, and long-baseline neutrino data~\cite{Fogli2007,Coloma2018,Holanda2020,Oliveira2016,BalieiroGomes2017,CoelhoMann2017,DeGouvea2020,DeGouvea2021,DeRomeri2023,BeraDeepthi2024,BeraDeepthi2025,AguilarESSnuSB2024,AielloKM3NeT2025,AbbasiIceCube2025}. The present work is therefore positioned as a quantum-information analysis of correlation measures under a dephasing map, rather than as a first study of neutrino decoherence itself.

Three benchmark experiments are particularly well suited for such a multidisciplinary analysis: KamLAND, MINOS, and Daya Bay. KamLAND observes the disappearance of electron antineutrinos ($\bar{\nu}_e$) from nuclear reactors at an effective baseline of about $180~\text{km}$, probing oscillations in the solar sector with parameters $\Delta m^2_{21}$ and $\theta_{12}$~\cite{Gando2013,kamland,Esteban2020,Abe2021}. MINOS uses a muon-neutrino beam produced at Fermilab and detects $\nu_\mu$ and $\bar{\nu}_\mu$ at a distance of $735~\text{km}$, primarily constraining the atmospheric sector parameters $\Delta m^2_{32}$ and $\theta_{23}$~\cite{Adamson2006,Adamson2008,Adamson2013,minos}. The Daya Bay experiment studies $\bar{\nu}_e$ disappearance over baselines of a few hundred to about two thousand meters, optimally probing the reactor sector and providing a high-precision determination of $\theta_{13}$ and the effective mass-squared difference $\Delta m^2_{ee}$~\cite{An2012,An2016,An2017,daya}. Taken together, these experiments span distinct regions of the oscillation parameter space and operate at different energies and baselines, making them ideal testbeds for exploring the behavior of quantum correlations in realistic neutrino propagation regimes.

Understanding the time-dependent behavior of quantum correlations—in particular EOF~\cite{Bennett1996,Wootters,Plenio2007}, QD~\cite{Ollivier2001,Henderson2001,Giorda2010,Modi2012}, and LQU~\cite{Girolami2013}—during neutrino oscillations~\cite{Blasone2009,blennow2010entanglement}
remains a central challenge. Previous works have already applied entanglement- and discord-based tools to oscillating neutrinos in a variety of settings, including two- and
three-flavor frameworks and field-theoretic treatments, and have clarified several aspects of neutrino quantumness and decoherence~\cite{Blasone2011,Alok2016,Dixit2019,Blasone2008,Wang2016}. In the present paper, we build on these analyses and focus on how complementary correlation measures behave in experimentally realistic two-flavor regimes, with explicit attention to environmental noise and open-system dynamics.

A closely related recent study by El Bouzaidi \textit{et al.}\cite{ElBouzaidi2025} considered the dynamics of quantum information resources in two-flavor neutrino oscillations for the same benchmark experiments, focusing on quantum coherence, entanglement, local quantum Fisher information, and Bell non-locality in the coherent oscillation setting. The present work is complementary in two respects: it concentrates on the triplet EOF--QD--LQU and, more importantly, it follows their degradation under a dephasing map that is explicitly connected below to the Lindblad damping parameter used in neutrino-decoherence phenomenology. Identifying which forms of quantumness are most robust against environmental noise is valuable both for fundamental tests of quantum theory and for possible applications in quantum metrology and information processing that exploit neutrino beams~\cite{giovannetti2011advances,pezze2018quantum}. Recently, quantum estimation theory has been applied directly to neutrino oscillation experiments to quantify ultimate precision bounds on the mixing parameters in terms of quantum and classical Fisher information~\cite{Frugiuele2026QET}. Our analysis complements this approach: rather than optimizing estimation strategies, we focus on nonclassical correlations such as EOF, QD, and, in particular, LQU, which also bounds the local quantum Fisher information and thus provides an experimentally oriented witness of metrological resources in two-flavor neutrino oscillations. Section~\ref{subsec:sensitivity} provides explicit derivatives with respect to the oscillation and dephasing parameters and clarifies the information content of the correlation measures relative to the ordinary transition probability.

For completeness, we also note that temporal non-classicality in neutrino oscillations has been investigated through Leggett--Garg inequalities, including two- and three-flavor treatments, quantum-field-theoretic analyses, non-standard interactions, and decoherence effects~\cite{WangMa2022LGI,Blasone2023LGI,ShafaqMehta2021,Naikoo2019LGI}. These works provide a complementary route to testing the coherence of neutrino oscillations, while the present manuscript focuses on EOF, QD, and LQU as correlation quantifiers.

In this work, we adopt a quantum-information perspective based on the triplet of entanglement of formation (EOF), quantum discord (QD), and local quantum uncertainty (LQU). Restricting ourselves to two-flavor oscillations allows us to derive simple closed expressions for these quantities directly in terms of the survival and transition probabilities, using parameter sets corresponding to the KamLAND, MINOS, and Daya Bay experiments. Within the same framework, we introduce a minimal dephasing channel that
models decoherence along the neutrino path and use it to track how the three measures respond to controlled suppression of coherence in realistic propagation scenarios. To facilitate comparison with the standard neutrino-decoherence literature, the dephasing amplitude is written as \(1-\gamma_{ij}(L,E)=\exp[-\Gamma_{ij}(E)L]\), where \(\Gamma_{ij}\) is the Lindblad damping parameter constrained by oscillation data.

In particular, we show that in the effective two-qubit description, the LQU reduces to the squared concurrence, thereby providing a direct operational link between skew-information-based quantifiers and standard entanglement measures. Because LQU also bounds the local quantum Fisher information, this connects neutrino oscillations to ultimate limits on parameter-estimation precision in noisy relativistic systems. Our analysis thus offers a compact, experimentally oriented quantum-information viewpoint on two-flavor neutrino oscillations—rather than a claim of absolute novelty—and identifies concrete regimes in which neutrinos behave as robust carriers of quantum correlations
in open quantum environments.	

\section{Neutrino oscillation analysis via bipartition mapping} 

The three neutrino flavor states electron (\(\nu_e\)), muon (\(\nu_\mu\)), and tau (\(\nu_\tau\)) do not align with mass eigenstates \cite{Pontecorvo1957}. Each flavor eigenstate constitutes a coherent linear superposition of the mass eigenstates \(\nu_1\), \(\nu_2\), and \(\nu_3\). This relationship is mathematically expressed through the lepton mixing matrix \cite{Maki1962}:
	\begin{equation}
		\ket{\nu_\alpha} = \sum_{i=1}^3 U_{\alpha i} \ket{\nu_i}, \quad \alpha \in \{e, \mu, \tau\}
		\label{eq:superposition}
	\end{equation}
In this framework, \(\ket{\nu_\alpha}\) denotes a flavor eigenstate, while \(\ket{\nu_i}\) represents a mass eigenstate with \(i = 1,2,3\). The coefficients \(U_{\alpha i}\) are the components of the Pontecorvo-Maki-Nakagawa-Sakata (PMNS) mixing matrix. This matrix is characterized by three mixing angles, \(\theta_{12}\), \(\theta_{13}\), \(\theta_{23}\), together with a complex phase \(\delta_{\mathrm{CP}}\), which accounts for possible CP violation in the lepton sector \cite{Bilenky2010,Tanabashi2018}.
The conventional way of expressing the PMNS matrix is through its standard parameterization, defined by \cite{Bilenky2010}:
\begin{widetext}
\begin{equation}
U =
\begin{bmatrix}
				c_{12}c_{13} & s_{12}c_{13} & s_{13}e^{-\mathrm{i}\delta_{\mathrm{CP}}} \\
				-s_{12}c_{23} - c_{12}s_{23}s_{13}e^{\mathrm{i}\delta_{\mathrm{CP}}} & c_{12}c_{23} - s_{12}s_{23}s_{13}e^{\mathrm{i}\delta_{\mathrm{CP}}} & s_{23}c_{13} \\
				s_{12}s_{23} - c_{12}c_{23}s_{13}e^{\mathrm{i}\delta_{\mathrm{CP}}} & -c_{12}s_{23} - s_{12}c_{23}s_{13}e^{\mathrm{i}\delta_{\mathrm{CP}}} & c_{23}c_{13}
			\end{bmatrix}
\label{eq:pmns_matrix}
\end{equation}
\end{widetext}
    
	\noindent in which \(c_{ij} \equiv \cos\theta_{ij}\) and \(s_{ij} \equiv \sin\theta_{ij}\) the angles for mixing  \(\theta_{12}\), \(\theta_{13}\), and \(\theta_{23}\).\\
	For two-flavor approximation, the full PMNS matrix (Eq.~\ref{eq:pmns_matrix}) simplifies to a \(2 \times 2\) rotation \cite{Bilenky2010}:
	\begin{equation}
		\begin{pmatrix}
			\nu_\alpha \\ \nu_\beta
		\end{pmatrix}
		=
		\begin{pmatrix}
			\cos\theta & \sin\theta \\
			-\sin\theta & \cos\theta
		\end{pmatrix}
		\begin{pmatrix}
			\nu_i \\ \nu_j
		\end{pmatrix}
		\label{eq:rotation}
	\end{equation}
	The analysis below employs the standard effective two-flavor approximation, appropriate when a single mass-squared splitting dominates the oscillation channel under consideration and matter effects can be neglected or absorbed into effective parameters over the relevant energy-baseline window. In this sense, the KamLAND, MINOS, and Daya Bay cases are used here as representative two-flavor regimes for comparing quantum-correlation measures, rather than as a substitute for a full three-flavor global-fit analysis, where \(\alpha, \beta\) denote distinct flavor indices (\(e, \mu, \tau\)), and \(i, j\) label mass eigenstates.\\

	The limitations of this projection should also be kept explicit. In full three-flavor phenomenology the three benchmark experiments receive corrections governed by the small quantities \(s_{13}^{2}\) and \(\alpha\equiv\Delta m_{21}^{2}/\Delta m_{31}^{2}\), by the solar phase in short-baseline reactor experiments, and by matter-induced shifts of the effective mixing parameters in accelerator experiments. For example, in the KamLAND regime, the leading three-flavor survival probability may be written approximately as
	\begin{equation}
	P_{\bar e\bar e}^{3\nu}\simeq c_{13}^{4}\left[1-\sin^{2}(2\theta_{12})\sin^{2}\Delta_{21}\right]+s_{13}^{4},
	\end{equation}
	so that the dominant two-flavor solar term is mainly rescaled by \(c_{13}^{4}\) with an additional small \(s_{13}^{4}\) contribution. In Daya Bay, the leading term is driven by \(\theta_{13}\) and \(\Delta m_{ee}^{2}\), while the solar correction is proportional to \(c_{13}^{4}\sin^{2}(2\theta_{12})\sin^{2}\Delta_{21}\) and is small at kilometer-scale baselines. In MINOS, the dominant behavior is controlled by \(\theta_{23}\) and the atmospheric mass splitting, with subleading corrections scaling with \(s_{13}^{2}\), \(\alpha\), and matter effects. These corrections are essential for precision global fits, but in the present quantum-information analysis, they mainly shift the effective phase and amplitude of the dominant transition probability. Because EOF, QD, and LQU are continuous functions of the transition probability and of the off-diagonal coherence terms, the qualitative hierarchy among KamLAND, MINOS, and Daya Bay and the coherence-loss interpretation of dephasing are not changed by these subleading corrections.\\
	
	Within this framework \cite{Pontecorvo1957,Maki1962,Giunti2007}, each mass eigenstate evolves temporally as:
	\begin{equation}
		\ket{\nu_i(t)} = e^{-\mathrm{i} E_i t} \ket{\nu_i},
		\label{eq:time_evolution}
	\end{equation}
	with \(E_i\) representing the energy. Consequently, a flavor state evolves according to:
	\begin{equation}
		\ket{\nu_\alpha(t)} = \sum_{i} e^{-\mathrm{i} E_i t} U_{\alpha i} \ket{\nu_i}.
		\label{eq:flavorevolution}
	\end{equation}
	Projecting Eq.~\eqref{eq:flavorevolution} onto the flavor basis yields:
	\begin{equation}
		\ket{\nu_\alpha(t)} = \tilde{a}_{\alpha\alpha}(t) \ket{\nu_\alpha} + \tilde{a}_{\alpha\beta}(t) \ket{\nu_\beta},
		\label{eq:projection}
	\end{equation}
	where normalization requires \(|\tilde{a}_{\alpha\alpha}(t)|^2 + |\tilde{a}_{\alpha\beta}(t)|^2 = 1\). These states can be rewritten in the composite two-qubit basis to yield:
	\begin{align}
		\ket{\nu_\alpha} &= \ket{\alpha} \otimes \ket{\beta} = \ket{\alpha\beta} \\
		\ket{\nu_\beta} &= \ket{\beta} \otimes \ket{\alpha} = \ket{\beta\alpha}
	\end{align}
	Eq.~\eqref{eq:flavorevolution} is re-expressed as:
	\begin{equation}
		\ket{\nu_\alpha(t)} = \tilde{a}_{\alpha\alpha}(t) \ket{\alpha\beta} + \tilde{a}_{\alpha\beta}(t) \ket{\beta\alpha}.
		\label{eq:entangledstate}
	\end{equation}
	The coefficients are explicitly given by:
	\begin{align}
		\tilde{a}_{\alpha\alpha}(t) &= \cos^2\theta \, e^{-\mathrm{i} E_i t} + \sin^2\theta \, e^{-\mathrm{i} E_j t}, \\
		\tilde{a}_{\alpha\beta}(t) &= \sin\theta \cos\theta \left( e^{-\mathrm{i} E_j t} - e^{-\mathrm{i} E_i t} \right).
		\label{eq:coefficients}
	\end{align}
	This state (Eq.\eqref{eq:entangledstate}) represents a single-particle entangled system \cite{Blasone2009}. The density matrix for the neutrino system in the specified basis $ \left\lbrace  \ket{\beta\beta} ,\ket{\beta\alpha}, \ket{\alpha\beta},\ket{\alpha\alpha}\right\rbrace  $ is given by:
		\begin{equation}
			\rho^{\alpha\beta}(t)=\ket{\nu_{\alpha}(t)}\bra{\nu_{\alpha}(t)},
	\end{equation}
which gives:
\begin{equation}
\begin{split}
\rho^{\alpha\beta}(t) &=  |\tilde{a}_{\alpha\beta}(t)|^2\ket{\beta\alpha}\bra{\beta\alpha} + \tilde{a}_{\alpha\alpha}(t)\tilde{a}^*_{\alpha\beta}(t)\ket{\beta\alpha}\bra{\alpha\beta}  \\
&\quad +\tilde{a}_{\alpha\beta}(t)\tilde{a}^*_{\alpha\alpha}(t)\ket{\alpha\beta}\bra{\beta\alpha} +|\tilde{a}_{\alpha\alpha}(t)|^2\ket{\alpha\beta}\bra{\alpha\beta}
\end{split}\label{eq:density_matrix}
\end{equation}
	The two-flavor survival and transition probabilities are then:
	\begin{align}
			|\tilde{a}_{\alpha\alpha}(t)|^2 =& c^4 + s^4 + 2s^2c^2\cos\left(\frac{\Delta m^2 t}{2E}\right) = P_{\alpha\alpha}\label{eq:probabilities}\\
			|\tilde{a}_{\alpha\beta}(t)|^2 =& 4s^2c^2\sin^2\left(\frac{\Delta m^2 t}{4E}\right) = P_{\alpha\beta}
	\end{align}
	Or coherence terms are written as:
    \begin{align}
       \tilde{a}_{\alpha\alpha}(t)\tilde{a}^*_{\alpha\beta}(t) 
		= s c &\left[ -\cos 2\theta + \cos 2\theta \cos\left(\frac{\Delta m^2 t}{2E}\right) \right.\notag\\&\left.+ \mathrm{i} \sin\left(\frac{\Delta m^2 t}{2E}\right) \right], 
    \end{align}
	\begin{align}
		\tilde{a}_{\alpha\beta}(t)\tilde{a}^*_{\alpha\alpha}(t) = s c &\left[ -\cos 2\theta + \cos 2\theta \cos\left(\frac{\Delta m^2 t}{2E}\right) \right.\notag\\&\left.- \mathrm{i} \sin\left(\frac{\Delta m^2 t}{2E}\right) \right],
	\end{align}
defined as $c = \cos(\theta)$ and $s = \sin(\theta)$.\\
	satisfying \(P_{\alpha\alpha} + P_{\alpha\beta} = 1\). Here, \(\Delta m^2 \equiv m_j^2 - m_i^2\). In the high-energy limit (\(m_i \ll E\)), the time \(t\) corresponds to the traveled distance \(L\), and the energy difference simplifies to \(E_j - E_i \simeq \Delta m^2/(2E)\) \cite{Giunti2007}.
	The survival and transition probabilities derived in Eq.~\ref{eq:probabilities} can be condensed into their canonical form \cite{Giunti2007,Bilenky2010}:
	\begin{align}
		P_{\alpha \to \alpha}(t) &= 1 - \sin^2(2\theta)\,\sin^2\!\left(\frac{\Delta m^2 t}{4E}\right) \\[4pt]
		P_{\alpha\to\beta}(t) &= \sin^2(2\theta)\,\sin^2\!\left(\frac{\Delta m^2 t}{4E}\right)
		\label{eq:standard_probs}
	\end{align}
	where \(\sin^2(2\theta) \equiv 4\sin^2\theta\cos^2\theta\) quantifies the oscillation amplitude. These satisfy \(P_{\alpha\alpha} + P_{\alpha\beta} = 1\) by construction.\\
The neutrino flavor transition probability exhibits a characteristic oscillatory behavior. 
The oscillation amplitude, given by \(\sin^2(2\theta)\), determines the maximum transition probability, 
bounded within the interval \([0,1]\). The expression governs the oscillation frequency 
\(\Delta m^2/4E\). This frequency defines a characteristic oscillation length, 
\(L_{\text{osc}} = \dfrac{4\pi E}{\Delta m^2}\), corresponding to the spatial period of the oscillations. 
In the ultra-relativistic limit, valid for neutrinos satisfying \(E \gg m_i\) 
and where the travel time \(t\) is approximately equal to the baseline distance \(L\) (\(t \simeq L\)), 
the phase difference driving the oscillations simplifies to \(\frac{\Delta m^2 t}{4E}\). 
This fundamental formalism, linking the amplitude, frequency, characteristic length, and phase, 
forms the cornerstone of all experimental analyses aimed at measuring 
the neutrino oscillation parameters \cite{Tanabashi2018, Giunti2007}.\\

 We initiate our investigation by characterizing quantum correlations in a two-qubit system emerging from two-flavor neutrino mixing. This framework provides a physically relevant approximation for three principal experimental configurations in neutrino oscillation physics, namely $\nu_\mu \rightarrow \nu_\tau$ transitions in atmospheric neutrino studies \cite{Fukuda1998,Ahmed2015}, $\nu_e \rightarrow \nu_\mu$ oscillations in reactor neutrino experiments \cite{An2017,Guo2020}, and $\nu_\mu \rightarrow \nu_e$ transitions in accelerator-based neutrino beams \cite{Abe2011,Adamson2013}. To quantify these correlations, we employ precision oscillation parameters determined by three benchmark experiments: Daya Bay \cite{An2017} for $\theta_{13}$ measurements via reactor $\bar{\nu}_e$ disappearance, KamLAND \cite{Gando2013} for $\Delta m_{21}^2$ determination through long-baseline reactor antineutrinos, and MINOS \cite{Adamson2013} for $\Delta m_{32}^2$ and $\theta_{23}$ constraints via accelerator $\nu_\mu$ disappearance. Each experiment probes distinct aspects of the oscillation parameter space, enabling a comprehensive quantum information-theoretic analysis across complementary energy and baseline regimes \cite{Bilenky2010,Giunti2007}.
	
\begin{table*}
	\centering
	\footnotesize
	\begin{tabular}{lcc}
		\hline
		Experiment & Oscillation Parameters & Baseline \& Energy Range \\
		\hline
		KamLAND & 
		
\mbox{
$\begin{aligned}
			\Delta m^2_{12} &= 7.49 \times 10^{-5} \text{ eV}^2 \\
			\tan^2 \theta_{12} &= 0.47
		\end{aligned}$
}
 &
		$L = 180\text{ km}, E = 2-10\text{ MeV}$ \\

		MINOS & 
		$\begin{aligned}
			\Delta m^2_{32} &= (2.32^{+0.12}_{-0.08}) \times 10^{-3} \text{ eV}^2 \\
			\sin^2 2\theta_{23} &= 0.95 \pm 0.035
		\end{aligned}$ &
		$L = 735\text{ km}, E = 0.5-50\text{ GeV}$ \\
		Daya Bay & 
		$\begin{aligned}
			\Delta m^2_{ee} &= (2.42^{+0.10}_{-0.11}) \times 10^{-3} \text{ eV}^2 \\
			\sin^2 2\theta_{13} &= 0.084 \pm 0.005
		\end{aligned}$ &
		$L = 364-1912\text{ m}, E = 1-8\text{ MeV}$ \\
		\hline
	\end{tabular}
	\caption{\footnotesize Summary of key parameters for the KamLAND, MINOS and Daya Bay experiments, including squared mass differences, mixing angles, baselines, and energy ranges.}
\end{table*}

 The KamLAND experiment, an intermediate-distance antineutrino disappearance study, analyzes electron antineutrinos from multiple nuclear reactors, with an effective baseline of approximately $L \simeq 180\ \text{km}$. Operating in the energy range $E \in [2,\ 10]\ \text{MeV}$, KamLAND confirmed key parameters of solar neutrino oscillations, notably 
\[
\Delta m^2_{21} = 7.49 \times 10^{-5}\ \text{eV}^2
\]
and the mixing angle $\theta_{12}$. This study was crucial for resolving the problem posed by solar neutrinos and for validating the Large Mixing Angle (LMA) scenario~\cite{kamland}.
\medskip
In comparison, the \textsc{MINOS} experiment is a long-distance study, conducted using a muon neutrino beam produced by an accelerator, with a baseline $L = 735\ \text{km}$. Operating at significantly higher energies, $E \in [0.5,\ 50]\ \text{GeV}$, MINOS imposed rigorous constraints on atmospheric neutrino oscillation parameters, in particular 
\[
\Delta m^2_{32} = (2.32^{+0.12}_{-0.08}) \times 10^{-3}\ \text{eV}^2
\]
and the mixing angle $\theta_{23}$. Its results were decisive in improving our understanding of dominant mixing in atmospheric neutrino oscillations~\cite{minos}.\\
Finally, the Daya Bay experiment constitutes a short-distance electron antineutrino disappearance study, examining antineutrinos emitted by nuclear reactors with baselines between $364\ \text{m}$ and $1912\ \text{m}$. In the energy regime $E \in [1,\ 8]\ \text{MeV}$, this experiment was instrumental in accurately determining the smallest mixing angle, $\theta_{13}$, along with the effective neutrino mass-squared difference.
\[
\Delta m^2_{ee} = (2.42^{+0.10}_{-0.11}) \times 10^{-3}\ \text{eV}^2 .
\]
These results have significantly enriched our understanding of subdominant mixing in neutrino oscillations~\cite{daya}.
\section{Quantum Information Resources}
Here, we systematically investigate three key quantifiers of quantum correlations: entanglement of formation, quantum discord and local quantum uncertainty, to track their dynamics in neutrino oscillations. These complementary tools capture different aspects of quantum correlations: EOF quantifies entanglement, QD detects non-classical correlations even in the absence of entanglement, and LQU measures sensitivity to local perturbations. We compare their baseline dependence for three representative experimental setups: KamLAND (solar sector, intermediate baseline), MINOS (atmospheric sector, long baseline), and Daya Bay (\(\theta_{13}\) angle, short baseline). This analysis highlights the influence of experimental parameters (\(\Delta m^2\), \(\theta\), \(L/E\)) on quantum signatures and reveals the convergences and differences among the three measures, demonstrating how neutrino oscillations provide a natural laboratory to explore quantum correlations at macroscopic scales.

\subsection{Quantum entanglement}
As a central characteristic of quantum mechanics, quantum entanglement reflects its inherently non-classical nature. It emerges when two or more systems exhibit correlations so profound that the quantum state of one cannot be described independently of the others, even when separated by large distances. This phenomenon, initially revealed through the EPR paradox \cite{Einstein1935} and experimentally verified via Bell's inequalities \cite{Bell1964}, challenges classical notions of locality and physical realism.
A particularly insightful demonstration of these non-classical effects occurs in neutrino oscillations. These elementary particles, abundantly produced in nuclear processes (e.g., within the Sun, Earth's atmosphere, or nuclear reactors), can transition between flavors (electron, muon, or tau) during propagation. This behavior comes directly from quantum superposition and the mixing of mass eigenstates \cite{Pontecorvo1957,Maki1962}.
Investigating entanglement in neutrino oscillations provides a distinctive framework for probing quantum features associated with flavor mixing and, potentially, departures from standard coherent propagation. When a flavor state is generated as a coherent mixture of mass states, its unitary evolution can be interpreted as a form of entanglement between mass modes \cite{Blasone2009,Blasone2014}. The entanglement of formation $E_F(\rho^{\alpha\beta}(t))$ is widely used to quantify entanglement in systems described by the reduced density matrix $\rho^{\alpha\beta}(t)$. This metric, closely associated with concurrence \cite{Wootters}, captures the minimal quantum resources essential to prepare the entangled state.\\
The formation-based entanglement measure  is typically computed via the convex roof extension method, defined as:
\begin{equation}
	E_F(\rho^{\alpha\beta}(t)) = \min_{\{p_j, \ket{\varphi_j}^{\alpha\beta}\}} \sum_j p_j E_F(\ket{\varphi_j}^{\alpha\beta})
\end{equation}
where the minimization ranges over all possible pure-state decompositions of $\rho^{\alpha\beta}(t) = \sum_j p_j \ket{\varphi_j}^{\alpha\beta}\bra{\varphi_j}$ satisfying $p_j \geq 0$ and $\sum_j p_j = 1$.

For a mixed two-qubit state $\rho^{\alpha\beta}(t)$, Wootters \cite{Wootters} derived an analytical expression:
\begin{equation}
	E_F(\rho^{\alpha\beta}(t)) = h\left(\frac{1}{2}\left( 1 + \sqrt{1 - |C(\rho^{\alpha\beta}(t))|^2}\right) \right)
\end{equation}
where $h(x) = -x\log_2 x - (1-x)\log_2(1-x)$ denotes the binary entropy and $C(\rho_{AB}) = \max\left\{0,\sqrt{\lambda_1} - \sqrt{\lambda_2} - \sqrt{\lambda_3} - \sqrt{\lambda_4} \right\}$ represents the density matrix concurrence. Here, $\lambda_i$ are the eigenvalues of $\rho_{AB}\tilde{\rho}_{AB}$ sorted in descending order, with $\tilde{\rho}^{\alpha\beta} = (\sigma_y \otimes \sigma_y) (\rho^{\alpha\beta})^* (\sigma_y \otimes \sigma_y)$.\\
Concurrence quantifies the resource cost for highly entangled states, where $C(\rho_{AB}) = 0$ signifies separability and $C(\rho_{AB}) = 1$ maximal entanglement.\\
Given the block-diagonal structure of $\rho^{\alpha\beta}(t)$, the analysis reduces to the subspace $\{|10\rangle, |01\rangle\}$. For pure states, concurrence simplifies to:
\begin{equation}
	\mathcal{C}(\rho^{\alpha\beta}(t)) = 2|\tilde{a}_{\alpha\alpha}(t)\tilde{a}_{\alpha\beta}^*(t)| = 2\sqrt{P_{\alpha\alpha}(t)P_{\alpha\beta}(t)}
\end{equation}
with $P_{\alpha\alpha}(t) = |\tilde{a}_{\alpha\alpha}(t)|^2$ and $P_{\alpha\beta}(t) = |\tilde{a}_{\alpha\beta}(t)|^2$.\\
This leads to the following expression for the entanglement of formation:
\begin{equation}
	E_F(\rho^{\alpha\beta}(t)) = h\left(\frac{1}{2}\left(1 + \sqrt{1 - 4P_{\alpha\alpha}(t)P_{\alpha\beta}(t)}\right) \right)
\end{equation}
Explicitly:

\begin{align}
E_F(\rho^{\alpha\beta}(t))
&= -\frac{1}{2}\left(1 + \sqrt{1 - 4P_{\alpha\alpha}(t)P_{\alpha\beta}(t)}\right) \notag\\
&\quad \times \log_2\left[ \frac{1}{2}\left(1 + \sqrt{1 - 4P_{\alpha\alpha}(t)P_{\alpha\beta}(t)}\right) \right] \notag\\
&\quad -\frac{1}{2}\left(1 - \sqrt{1 - 4P_{\alpha\alpha}(t)P_{\alpha\beta}(t)}\right) \notag\\
&\quad \times \log_2\left[ \frac{1}{2}\left(1 - \sqrt{1 - 4P_{\alpha\alpha}(t)P_{\alpha\beta}(t)}\right) \right]
\label{eof}
\end{align}
The presence of nonlinearity in the expression \ref{eof} is characteristic of purely quantum behavior, including superposition. Similarly, the expression \ref{eof} describes the baseline dependence of entanglement in an oscillating neutrino system. Quantifies how non-classical quantum correlations emerge from flavor superpositions, 
through a dynamic interplay between quantum coherence 
(\(P_{\alpha\alpha} \leftrightarrow P_{\alpha\beta}\)) and disentanglement. Its behavior oscillates in phase with flavor transitions, establishing it as a dynamic entanglement witness for probing fundamental physics.

\subsection{Quantum discord}
Quantum discord evaluates correlations beyond classical descriptions in bipartite systems that persist when entanglement vanishes \cite{Ollivier2001}. Defining operationally as the difference between quantum extensions of mutual information \cite{Ollivier2001,Henderson2001}, it captures correlations detectable only through coherence that disrupts local measurements \cite{Zurek2003}. QD has been widely investigated in the study of two-flavor neutrino oscillations using different formalisms \cite{Ali2010,Blasone2014, slaoui}. In the present work, we adopt an alternative approach to quantifying QD, originally proposed by Wang et al. \cite{Wang2014}. The quantum discord \(Q(\rho^{\alpha\beta}(t))\) associated with a two-qubit density matrix \(\rho^{\alpha\beta}(t)\) is expressed as:
\begin{equation}
	Q(\rho^{\alpha\beta}(t)) := I(\rho^{\alpha\beta}(t)) - J(\rho^{\alpha\beta}(t)) \label{eq:DQ}
\end{equation}
where the total quantum correlations are quantified by the quantum mutual information:
\begin{equation}
	I(\rho^{\alpha\beta}(t)) := S(\rho_\alpha(t)) + S(\rho_\beta(t)) - S(\rho^{\alpha\beta}(t)),
\end{equation}
and $J(\rho^{\alpha\beta}(t))$ quantifies the classical correlations as follows:
\begin{equation}
	J(\rho^{\alpha\beta}(t)) = \max_{\{ \Pi_\beta^j \}} \left[ S(\rho_\beta) - \sum_j p_{\beta,j} S(\rho_{\beta|j}) \right]. \label{eq:Jclassique}
\end{equation}
The maximization is carried out over all positive operator-valued measures (POVMs) $\{ \Pi_\beta^j \}$ acting on subsystem $B$, satisfying:
\begin{equation}
    \sum_j \Pi_\beta^{j\dagger} \Pi_\beta^j = I
\end{equation}
The von Neumann entropy is given by $S(\rho) = - \mathrm{Tr}(\rho \log_2 \rho)$. The components in \eqref{eq:Jclassique} are defined as:
\begin{align}
	p_{\beta,j} &= \mathrm{Tr}_{\alpha\beta} \left[ (I \otimes \Pi_\beta^j) \rho^{\alpha\beta}(t) (I \otimes \Pi_\beta^{j\dagger}) \right], \\
	\rho_{\beta|j} &= \mathrm{Tr}_\alpha \left[ (I \otimes \Pi_\beta^j) \rho^{\alpha\beta}(t) (I \otimes \Pi_\beta^{j\dagger}) \right],
\end{align}
where $p_{\beta,j}$ represents the probability of outcome $j$, and $\rho_{\beta|j}$ is the conditional state of subsystem $\beta$ for this outcome. The fundamental concept in the calculation of quantum discord (QD) is to quantify the information that cannot be accessed via local measurements, extracting partial information about subsystem $\alpha$ from subsystem $\beta$ without disturbing $\alpha$. The most challenging part of the process is deriving an analytical expression for classical correlations, which requires optimization over all possible local measurements. As a result, closed-form expressions for QD are only available for specific two-qubit systems \cite{Dakic2010,Luo2008}. For the bipartite system in question, the analytical expression for QD can be derived using the method developed by C. Z. Wang et al. \cite{Wang2014}. In this framework, the two-flavor oscillation state is represented by an X-shaped density matrix:
\begin{equation}
\begin{split}
\rho^{\alpha\beta}(t) &=  \rho_{22}\ket{\beta\alpha}\bra{\beta\alpha} + \rho_{23}\ket{\beta\alpha}\bra{\alpha\beta}  \\
&\quad +\rho_{23}^*\ket{\alpha\beta}\bra{\beta\alpha} +\rho_{33}\ket{\alpha\beta}\bra{\alpha\beta}
\end{split}\label{eq:Xstate}
\end{equation}
with $\rho_{22} = |\tilde{a}_{\alpha\beta}(t)|^{2}$, $\rho_{33} = |\tilde{a}_{\alpha\alpha}(t)|^{2}$, 
$\rho_{23} = \tilde{a}_{\alpha\alpha}(t)\tilde{a}_{\alpha\beta}(t)^{*}$, and $\rho_{32} = \tilde{a}_{\alpha\beta}(t)\tilde{a}_{\alpha\alpha}(t)^{*}$, 
the eigenvalues of $\rho^{\alpha\beta}(t)$ are:

\begin{align}
	\lambda_{\pm} &= \frac{1}{2} \left[ \rho_{22} + \rho_{33} \pm \sqrt{(\rho_{22} - \rho_{33})^2 + 4|\rho_{23}|^2} \right].
\end{align}

The reduced matrix entropies $\rho_A$ and $\rho_B$ are given by:

\begin{align}
	S(\rho_\alpha) &= - \rho_{22} \log_2 \rho_{22} - \rho_{33} \log_2 \rho_{33}, \label{eq:SA} \\
	S(\rho_\beta) &= - \rho_{33} \log_2 \rho_{33} - \rho_{22} \log_2 \rho_{22}. \label{eq:SB}
\end{align}

To minimize the classical correlations in Eq.~\eqref{eq:Jclassique}, we consider a complete set of projective measurements on subsystem $B$:

\[
\Pi_\beta^j = |\beta_j\rangle \langle \beta_j|, \quad j = 1, 2,
\]
where:

\begin{align}
	|\beta_1\rangle &= \cos\left( \frac{\theta}{2} \right)|1\rangle + e^{\mathrm{i}\varphi} \sin\left( \frac{\theta}{2} \right)|0\rangle, \\
	|\beta_2\rangle &= \sin\left( \frac{\theta}{2} \right)|1\rangle - e^{\mathrm{i}\varphi} \cos\left( \frac{\theta}{2} \right)|0\rangle,
\end{align}
with $0 \leq \theta \leq \frac{\pi}{2}$ and $0 \leq \varphi < 2\pi$. The probability $p_{B,j}$ for outcome $j$ and the eigenvalues of the post-measurement state $\rho_{B,j}$ are:

\begin{equation}
	p_{\beta,j} = \frac{1}{2} \left[1 + (-1)^j \cos\theta \left(1 - 2\rho_{33}\right) \right] \label{eq:pbj}
\end{equation}

\begin{equation}
	\lambda_\pm(\rho_{\beta,j}) = \frac{1}{2} \left(1 \pm \frac{\sqrt{\Delta_j}}{p_{\beta,j}} \right) \label{eq:eigval}
\end{equation}
with
\begin{align}
    \Delta_j = &\frac{1}{4} \Big[ 1 - 2\rho_{33} + (-1)^j \cos\theta \Big]^2 + \sin^2\theta \left[ |\rho_{23}|^2 \right.\notag\\&\left.- 2|\rho_{23}| \sin(2\varphi + \delta) \right]. \label{eq:Delta}
\end{align}
The entropy of $\rho_{\beta,j}$ is calculated from its eigenvalues:

\begin{equation}
	S(\rho_{\beta,j}) = H(\lambda_+(\rho_{\beta,j})) \label{eq:entropy_Bj}
\end{equation}
where $H(x) = -x\log_2 x - (1 - x) \log_2(1 - x)$ is the binary Shannon entropy. The conditional entropy is given by:
\begin{equation}
	S(\rho_{\alpha|\beta}) = \sum_{j=1}^2 p_{\beta,j} S(\rho_{\beta,j}).\label{eq:SAB}
\end{equation}
Setting the partial derivatives of this conditional entropy with respect to $\theta$ and $\varphi$ to zero, we find that the minimum occurs at $\theta = \pi/2$ (where $p_{B,1} = p_{B,2}$ and $S(\rho_{B,1}) = S(\rho_{B,2})$). This gives the first conditional entropy extremum:

\begin{equation}
	\xi_1 = H\left( \frac{1 + \sqrt{(1 - 2\rho_{33})^2 + 4|\rho_{23}|^2}}{2} \right) \label{eq:chi1}
\end{equation}

The second extremum occurs at $\theta = 0$ (for arbitrary $\varphi$), giving:

\begin{equation}
	\xi_2 = -\sum_{i=\pm} \lambda_i \log_2 \lambda_i - H(\rho_{33}) \label{eq:chi2}
\end{equation}
The classical correlations for X-states (Eq.~\eqref{eq:Xstate}) are:
\begin{equation}
	J(\rho^{\alpha\beta}(t)) = \max(J_1, J_2) \label{eq:Jfinal}
\end{equation}
where 

\begin{equation}
	J_i = H(\rho_{22}) - \xi_i \label{eq:Ji}
\end{equation}

The analytic quantum discord is then obtained from Eq.~\eqref{eq:DQ} as:

\begin{equation}
	Q(\rho^{\alpha\beta}(t)) = \min(Q_1, Q_2) \label{eq:Qfinal}
\end{equation}
with

\begin{equation}
	Q_i = H(\rho_{33}) + \sum_{i=\pm} \lambda_i \log_2 \lambda_i + \xi_i \label{eq:Qi}
\end{equation}

For a two-qubit system, quantum discord simplifies to:

\begin{align}
	Q_1 &= H\left(\rho_{22}\right) + \lambda_{+} \log_2 \lambda_{+} + \lambda_{-} \log_2 \lambda_{-} \notag \\
	&\quad + H\left( \frac{1 + \sqrt{(1 - 2\rho_{33})^2 + 4|\rho_{23}|^2}}{2} \right) \\
	Q_2 &= 2|\rho_{23}|^2 \label{eq:Q2}
\end{align}

For X-states, the quantum discord expression condenses to~\cite{Wang2014}:

\begin{widetext}
\begin{align}
	Q(\rho^{\alpha\beta}(t)) &= \min \left[ H\left(\rho_{22}\right) + \lambda_{+} \log_2 \lambda_{+} + \lambda_{-} \log_2 \lambda_{-} \right. \notag \\
	&\left. \quad + H\left( \frac{1 + \sqrt{(1 - 2\rho_{33})^2 + 4|\rho_{23}|^2}}{2} \right), \; 2|\rho_{23}|^2 \right]
\end{align}
\end{widetext}

where $\lambda_{\pm}$ are the system eigenvalues:

\begin{align}
	\lambda_{\pm} = \frac{1}{2} \left(1 \pm \sqrt{1 - 4|\tilde{a}_{\alpha\beta}(t)\tilde{a}_{\alpha\alpha}(t)|^2} \right).
\end{align}

The quantum discord further simplifies to:

\begin{align}
	Q(\rho^{\alpha\beta}(t)) = \min\Big[& H\left(|\tilde{a}_{\alpha\beta}(t)|^2\right),\notag\\
	&2 |\tilde{a}_{\alpha\beta}(t)|^2\left(1 - |\tilde{a}_{\alpha\beta}(t)|^2\right) \Big].
\end{align}

Using the analytic method for X-states~\cite{Wang2014}, the quantum discord is compactly expressed as:

\begin{equation}
	Q(\rho^{\alpha\beta}(t)) = \min\left[ H\left(P_{\alpha\beta}\right), \; 2 P_{\alpha\beta} P_{\alpha\alpha} \right], \label{eq:final_QD}
\end{equation}
where \(H(x) = -x \log_{2} x - (1-x) \log_{2} (1-x)\) is the binary entropy function. This expression shows that quantum discord depends on the minimum of the transition probability entropy $H(P_{\alpha\beta})$ and the product $2P_{\alpha\beta}P_{\alpha\alpha}$, highlighting its sensitivity to quantum amplitude distributions in bipartite states.

\subsection{Local quantum uncertainty}
\label{subsec:lqu}

The local quantum uncertainty (LQU) is a discord-type measure based on the minimum Wigner--Yanase skew information generated by local observables. For a bipartite state \(\rho_{AB}\), it is defined as~\cite{Girolami2013}
\begin{equation}
\mathcal{U}(\rho_{AB})=\min_{K_A}\mathcal{I}(\rho_{AB},K_A\otimes I_B),
\end{equation}
where
\begin{equation}
\mathcal{I}(\rho,H)=-\frac{1}{2}\operatorname{Tr}\left([\sqrt{\rho},H]^2\right).
\end{equation}
For two-qubit states this quantity has the closed form
\begin{equation}
\mathcal{U}(\rho)=1-\max\{\lambda_1,\lambda_2,\lambda_3\},
\label{eq:LQU_closed_form}
\end{equation}
where \(\lambda_i\) are the eigenvalues of the \(3\times3\) matrix
\begin{equation}
W_{ij}=\operatorname{Tr}\left[\sqrt{\rho}\,(\sigma_i\otimes I_2)\sqrt{\rho}\,(\sigma_j\otimes I_2)\right],\qquad i,j=1,2,3.
\label{eq:LQU_W_matrix}
\end{equation}
Applying this formula to the pure two-flavor state in Eq.~\eqref{eq:entangledstate} gives
\begin{equation}
\mathcal{U}(\rho^{\alpha\beta}(t))=4|\tilde a_{\alpha\alpha}(t)|^2|\tilde a_{\alpha\beta}(t)|^2=4P_{\alpha\alpha}(t)P_{\alpha\beta}(t).
\label{eq:LQU_final}
\end{equation}
Equivalently, for the pure two-qubit state considered here,
\begin{equation}
\mathcal{U}(\rho^{\alpha\beta}(t))=\mathcal{C}^2(\rho^{\alpha\beta}(t)),
\end{equation}
so LQU tracks the squared concurrence and therefore provides a compact skew-information witness of the same entanglement generated by the flavor superposition. The intermediate algebra leading to Eq.~\eqref{eq:LQU_final} is moved to Appendix~\ref{app:lqu_derivation} in order to keep the main text focused on the physical implications.

\subsection{Sensitivity of the correlation measures to oscillation and dephasing parameters}
\label{subsec:sensitivity}

To make the connection with parameter sensitivity explicit, let
\begin{align}
P &\equiv P_{\alpha\beta}=\sin^2(2\theta)\sin^2\Phi,\notag\\
\Phi&=1.267\,\frac{\Delta m^2(\mathrm{eV}^2)L(\mathrm{km})}{E(\mathrm{GeV})}.
\label{eq:sensitivity_probability}
\end{align}
so that all pure-state quantifiers used above are functions of the single transition probability: \(\mathcal{U}=4P(1-P)\), \(Q=2P(1-P)\) for the relevant X-state branch, \(\mathcal{C}=2\sqrt{P(1-P)}\), and \(E_F=h[(1+\sqrt{1-\mathcal{C}^2})/2]\). For any oscillation parameter \(x\in\{\theta,\Delta m^2,L,E\}\), the chain rule gives
\begin{equation}
\frac{\partial M}{\partial x}=\frac{dM}{dP}\frac{\partial P}{\partial x},
\label{eq:chain_sensitivity}
\end{equation}
where
\begin{align}
\frac{\partial P}{\partial \theta}&=2\sin(4\theta)\sin^2\Phi, \notag\\
\frac{\partial P}{\partial (\Delta m^2)}&=\sin^2(2\theta)\sin(2\Phi)\,\frac{1.267L}{E},\notag\\
\frac{\partial P}{\partial L}&=\sin^2(2\theta)\sin(2\Phi)\,\frac{1.267\Delta m^2}{E}, \notag\\
\frac{\partial P}{\partial E}&=-\sin^2(2\theta)\sin(2\Phi)\,\frac{1.267\Delta m^2L}{E^2}.
\end{align}
The corresponding probability-space response factors are
\begin{align}
\frac{d\mathcal{U}}{dP}&=4(1-2P),\notag\\
\frac{dQ}{dP}&=2(1-2P),\notag\\
\frac{d\mathcal{C}}{dP}&=\frac{1-2P}{\sqrt{P(1-P)}}.
\label{eq:response_factors}
\end{align}
For EOF one may use
\begin{align}
\frac{dE_F}{dP}&=\frac{dE_F}{d\mathcal{C}}\frac{d\mathcal{C}}{dP},\notag\\
\frac{dE_F}{d\mathcal{C}}&=-\frac{\mathcal{C}}{2\sqrt{1-\mathcal{C}^2}}
\log_2\!\left(
\frac{1-\sqrt{1-\mathcal{C}^2}}{1+\sqrt{1-\mathcal{C}^2}}
\right).
\label{eq:eof_sensitivity}
\end{align}
Equations~\eqref{eq:chain_sensitivity}--\eqref{eq:eof_sensitivity} show that, in an ideal unitary two-flavor problem, the correlation quantifiers do not contain independent information on \(\theta\) or \(\Delta m^2\) beyond the full probability \(P_{\alpha\beta}\); they reweight the same oscillation information and emphasize different operating regions of the oscillation cycle. Their additional diagnostic value appears when coherence itself is a parameter. For the dephased state,
\begin{align}
|\rho_{23}|&=(1-\gamma)\sqrt{P(1-P)},\notag\\
\frac{\partial P}{\partial\gamma}&=0.
\label{eq:probability_gamma_zero}
\end{align}
whereas
\begin{align}
\frac{\partial\mathcal{C}_D}{\partial\gamma}&=-2\sqrt{P(1-P)},\notag\\
\frac{\partial Q_D}{\partial\gamma}&=-4(1-\gamma)P(1-P).
\label{eq:dephasing_sensitivity}
\end{align}
for the concurrence and the relevant QD branch. At the maximum-admixture point \(P=1/2\), Eq.~\eqref{eq:dephasing_sensitivity} gives \(|\partial\mathcal{C}_D/\partial\gamma|=1\) and, at \(\gamma=0\), \(|\partial Q_D/\partial\gamma|=1\), while the flavor probability remains insensitive to pure dephasing in this channel. This is the quantitative sense in which the correlation measures provide information complementary to the oscillation probability: they are direct probes of off-diagonal coherence loss, not merely of population transfer.

\begin{table*}
\centering

\caption{Local sensitivity factors of the main correlation measures. Here \(M_x\equiv |\partial M/\partial x|\) and \(P_x\equiv |\partial P/\partial x|\) for \(x=\theta,\Delta m^2,L,E\).}
\label{tab:sensitivity_factors}
\begin{tabular}{lcc}
\hline
Observable & Sensitivity to oscillation parameters & Sensitivity to pure dephasing \\
\hline
Transition probability \(P\) & \(P_x\) & \(0\) \\
LQU \(\mathcal{U}=4P(1-P)\) & \(|4(1-2P)|P_x\) & through Eq.~\eqref{eq:LQU_final} \\
QD \(Q_D\) & \(|2(1-2P)|P_x\) in the unitary branch & \(4(1-\gamma)P(1-P)\) \\
Concurrence \(\mathcal{C}_D\) & \(\left|\frac{1-2P}{\sqrt{P(1-P)}}\right|P_x\) & \(2\sqrt{P(1-P)}\) \\
EOF \(E_F\) & \(\left|\frac{dE_F}{d\mathcal{C}}\frac{d\mathcal{C}}{dP}\right|P_x\) & through \(\mathcal{C}_D=2(1-\gamma)\sqrt{P(1-P)}\) \\
\hline
\end{tabular}
\end{table*}

\begin{figure*}[t]
\centering
\begin{minipage}[b]{0.32\textwidth}
\centering
\includegraphics[width=\linewidth]{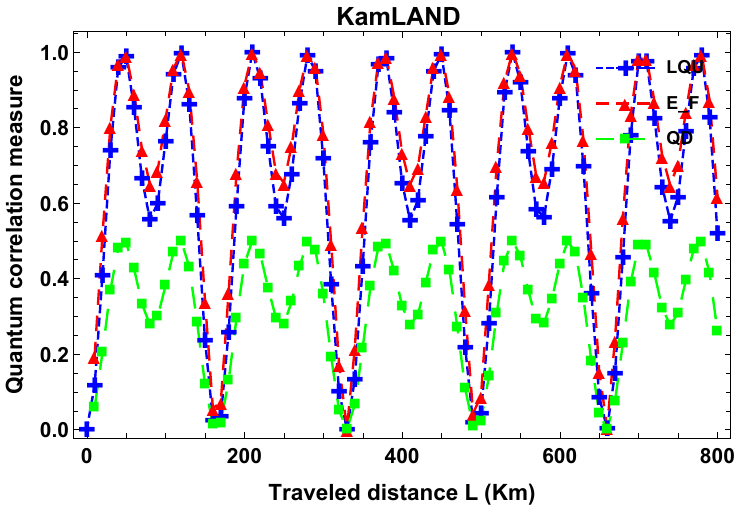}\par
(a)
\end{minipage}
\hfill
\begin{minipage}[b]{0.32\textwidth}
\centering
\includegraphics[width=\linewidth]{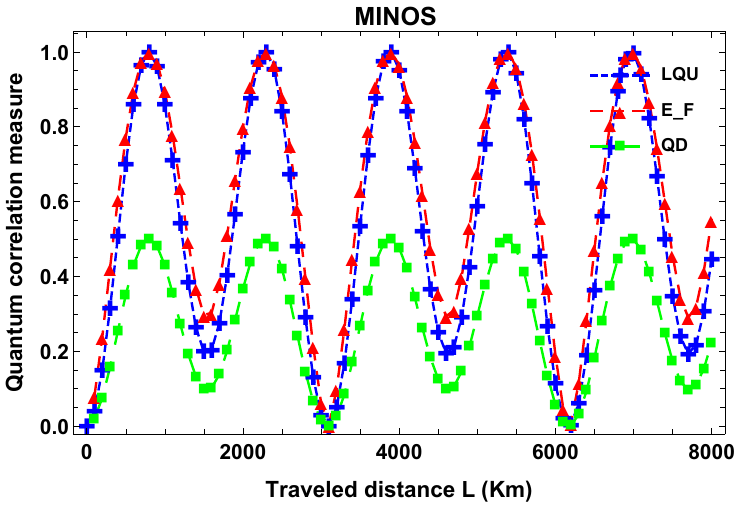}\par
(b)
\end{minipage}
\hfill
\begin{minipage}[b]{0.32\textwidth}
\centering
\includegraphics[width=\linewidth]{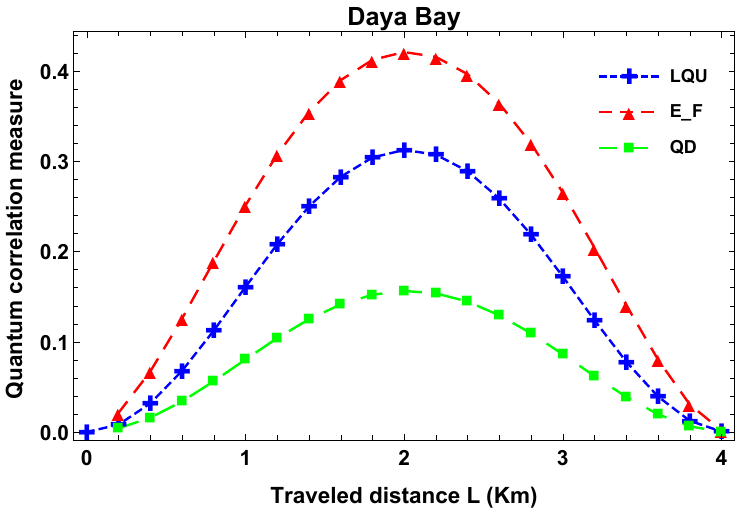}\par
(c)
\end{minipage}
\caption{Plots of Local Quantum Uncertainty, Quantum Discord, and Entanglement of Formation as functions of the traveled distance \(L\) in km for three experiments: (a) KamLAND, (b) MINOS, and (c) Daya Bay. The ultra-relativistic relation \(t\simeq L\) is used only to translate the oscillation phase into a baseline variable.}
\label{Fig45}
\end{figure*}

Figure~\ref{Fig45} presents the baseline dependence of bipartite quantum correlations in two-flavor neutrino oscillations, computed using experimental parameters from KamLAND, MINOS, and Daya Bay. Each correlation measure is plotted as a function of the traveled distance \(L\), revealing how the mass-squared difference \(\Delta m^2\), neutrino energy \(E\), and mixing angle \(\theta\) control the quantum signatures across different experimental regimes.\par

The KamLAND experiment [Fig.\ref{Fig45}(a)] probes $\bar{\nu}_e \to \bar{\nu}_\mu$ oscillations governed by solar sector parameters: $\Delta m^2_{21} = 7.49 \times 10^{-5}\text{eV}^2$ and a non-maximal mixing angle $\theta_{12} \approx 33.5^\circ$~\cite{Gando2013}. The LQU and Entanglement of Formation ($E_F$) exhibit synchronized oscillations, reaching peak values ($\mathcal{U} \approx 0.95$, $E_F \approx 0.95$) at points where the flavor composition achieves maximum admixture—specifically, when the quantum state contains comparable superpositions of $\ket{\nu_e}$ and $\ket{\nu_\mu}$~\cite{Blasone2009}. Critically, these maxima occur at intermediate baseline distances, not at the oscillation maxima where $P_{e\to\mu}$ is largest. Because $\theta_{12}$ is non-maximal, the flavor conversion remains incomplete even when $P_{e\to\mu}$ peaks, preventing the system from reaching a pure single-flavor state. Consequently, quantum correlation measures never fully vanish, maintaining non-zero minima throughout the oscillation cycle. This behavior validates the relation $\mathcal{U} = \mathcal{C}^2$ for pure bipartite states~\cite{Girolami2013}. Quantum Discord (QD), though exhibiting lower amplitude ($\sim 0.45$), persists across all baseline distances, capturing non-classical correlations beyond entanglement\cite{Banerjee2015}. The observed dynamics, modulated by $\sin^2(\Delta m^2_{21} L / 4E)$ with $L = 180\text{km}$ and $E \sim 4\text{MeV}$, demonstrate that intermediate-baseline reactor experiments provide robust testbeds for probing quantum coherence in flavor mixing.

The MINOS experiment Fig.\ref{Fig45}(b) investigates $\nu_\mu \to \nu_\tau$ oscillations over $L = 735~\text{km}$, characterized by atmospheric parameters $\Delta m^2_{32} \approx 2.4 \times 10^{-3}\text{eV}^2$ and near-maximal mixing $\theta_{23} \simeq 45^\circ$ ($\sin^2 2\theta_{23} \simeq 1$)\cite{Adamson2013}. Both LQU and $E_F$ reach their maximum values ($\mathcal{U} \approx 1$, $E_F \approx 1$) at baseline distances corresponding to maximal flavor admixture—that is, when the neutrino state forms an equal superposition of $\ket{\nu_\mu}$ and $\ket{\nu_\tau}$. This occurs at the midpoint between production and complete oscillation. Conversely, at distances where $P_{\mu\to\tau} \approx 1$ (near-complete conversion to $\nu_\tau$), the state approaches a pure single-flavor eigenstate, causing quantum correlation measures to drop toward their minima. This counterintuitive behavior arises because maximal mixing permits nearly total flavor conversion, leaving minimal admixture at oscillation peaks. The near-unity correlation values confirm $\mathcal{U} \approx \mathcal{C}^2$ and establish MINOS as an optimal configuration for generating strong bipartite entanglement\cite{Girolami2013}. QD maintains non-zero values ($\sim 0.4$) even when $E_F$ approaches zero, signaling persistent quantum correlations independent of entanglement\cite{Modi2012}. The oscillation phase $\Delta m^2_{32} L / 4E$ with $E \sim 3\text{GeV}$ ensures high sensitivity to atmospheric mass splitting, positioning MINOS as a premier platform for studying entanglement dynamics in long-baseline neutrino physics.

The Daya Bay experiment Fig.\ref{Fig45}(c) measures $\bar{\nu}_e$ disappearance at short baselines ($L \sim 0.5$–$2\text{km}$), dominated by the smallest mixing angle $\theta_{13} \approx 8.5^\circ$ and reactor mass splitting $\Delta m^2_{ee} \approx 2.5 \times 10^{-3}\text{eV}^2$\cite{An2017}. Here, $E_F$ and LQU exhibit substantially weaker oscillations ($\sim 0.45$) compared to KamLAND and MINOS. The small mixing angle implies that flavor conversion remains modest even at oscillation maxima, resulting in perpetual flavor admixture. Consequently, the maxima in quantum correlations coincide with oscillation peaks—a pattern opposite to that observed in MINOS. QD persists at low but non-zero levels ($\sim 0.15$), confirming that non-classical signatures survive even in weakly-mixing regimes\cite{Blasone2009,Modi2012}. Despite reduced amplitudes, Daya Bay demonstrates that quantum information-theoretic measures remain detectable and physically meaningful across all oscillation sectors.

The comparison across experiments reveals that the quantum correlation strength is primarily determined by the magnitude of the mixing angle, not the baseline length alone. MINOS achieves near-maximal correlations due to $\theta_{23} \simeq 45^\circ$, allowing strong flavor admixture. KamLAND exhibits intermediate values reflecting $\theta_{12} \approx 33.5^\circ$, while Daya Bay shows the weakest correlations corresponding to $\theta_{13} \approx 8.5^\circ$. The baseline distance and energy optimize the experimental sensitivity to specific values $\Delta m^2$ through the resonance condition $\Delta m^2 L / 4E \approx \pi/2$, but the mixing angle fundamentally governs the maximum achievable quantumness. These findings establish neutrino oscillations as a natural arena for exploring quantum correlations in relativistic open systems, with potential applications in quantum metrology and fundamental tests of quantum mechanics at macroscopic scales.
\section{Dynamics of quantum information resources under a dephasing channel}
Within the framework of neutrino oscillations, the quantum coherence between mass eigenstates plays a pivotal role in sustaining the interference responsible for flavor transitions \cite{Bilenky2010}. 
In realistic conditions, however, this coherence may degrade due to environmental interactions or long-baseline propagation effects, leading to decoherence phenomena \cite{Blasone2009}. 
	A widely used and physically relevant description of this type of noise is provided by the phase-damping channel, which selectively suppresses the off-diagonal components of the density matrix carriers of phase information while leaving the diagonal populations intact \cite{NielsenChuang, safae}. When applied to a two-flavor oscillation scenario, this channel enables the modeling of a progressive reduction in quantum correlations without altering the instantaneous detection probabilities in each flavor \cite{Bilenky2010,Blasone2009}.\par 

	Investigating this mechanism, particularly through the time evolution of indicators such as the entanglement of formation, quantum discord, or local quantum uncertainty, offers an effective approach to quantify the environmental impact on neutrino quantum dynamics and to assess the resilience of quantum resources under dissipative noise~\cite{Girolami2013,Blasone2009,Bilenky2010}.
	The phase-damping channel represents a quantum noise model that acts solely on the coherence terms (off-diagonal entries) of the density matrix, while preserving the populations (diagonal entries).
	To formalize such decoherence effects, we denote by \(\mathcal{\chi}\) the quantum map that transforms the physical state represented by the density operator \(\rho\) into another state \(\rho'\), parameterized by a decoherence probability \(p\):
	\begin{equation}
		\rho' = \mathcal{\chi}(\rho) = \sum_m K_m \rho K_m^\dagger 
	\end{equation}
	Here, the \(K_m\) are Kraus operators fulfilling the completeness relation:
	\begin{equation}
		\sum_m K_m^\dagger K_m = \mathbb{I}
	\end{equation}
	Such quantum maps modeling environmental effects are collectively referred to as quantum channels~\cite{ref47, ref48}. \\

	A practical and illustrative case is the phase-damping channel, which serves as a prototypical example of decoherence in physically relevant settings.
	Its action on the first qubit is expressed as:
	\begin{equation}
		\rho' = \left(1 - \frac{\gamma}{2}\right) \rho + \frac{\gamma}{2} \left(\sigma_3 \otimes I\right) \rho \left(\sigma_3 \otimes I\right)
	\end{equation}
	with effective dephasing strength \(\gamma_{ij}(L,E)=1-e^{-\Gamma_{ij}(E)L}\in[0,1]\) and \(\sigma_3 = \begin{pmatrix} 1 & 0 \\ 0 & -1 \end{pmatrix}\). This choice connects the finite Kraus map directly to the Lindblad damping parameter constrained in neutrino-oscillation analyses. In basis 
	\begin{equation}
	    \left\{ |\beta\beta\rangle, |\beta\alpha\rangle, |\alpha\beta\rangle, |\alpha\alpha\rangle \right\}
	\end{equation}
	the nonzero elements of \(\rho\) are:
	\begin{align}
		\rho_{22} &= |\tilde{a}_{\alpha\beta} (t)|^2 \\
		\rho_{23} &= \tilde{a}_{\alpha\alpha}(t)\tilde{a}_{\alpha\beta}^*(t) \\
		\rho_{32} &= \tilde{a}_{\alpha\beta}(t)\tilde{a}_{\alpha\alpha}^*(t) \\
		\rho_{33} &= |\tilde{a}_{\alpha\alpha} (t)|^2
	\end{align}
	Applying \(\sigma_3 \otimes I\) changes the sign of specific components according to their indices:
	\begin{equation}
		\sigma_3 \otimes I \, |\beta\alpha\rangle = +|\beta\alpha\rangle, \quad
		\sigma_3 \otimes I \, |\alpha\beta\rangle = -|\alpha\beta\rangle.
	\end{equation}
	From this, we find:
	\begin{align}
\rho'_{22} &= \rho_{22}, &
\rho'_{33} &= \rho_{33},\notag\\
\rho'_{23} &= (1 - \gamma)\rho_{23}, &
\rho'_{32} &= (1 - \gamma)\rho_{32}.
\end{align}
	Hence, the diagonal entries remain unchanged, whereas the coherence terms are attenuated by the factor \((1 - \gamma)\). 
	The resulting density matrix reads:
\begin{widetext}
\begin{equation}\label{rod}
\begin{split}
\rho_{\alpha\beta}^{D}(t) &= |\tilde{a}_{\alpha\beta}(t)|^2 \ket{\beta\alpha}\bra{\beta\alpha} + (1 - \gamma)\tilde{a}_{\alpha\alpha}(t)\tilde{a}^*_{\alpha\beta}(t) \ket{\beta\alpha}\bra{\alpha\beta} \\
&\quad + (1 - \gamma)\tilde{a}_{\alpha\beta}(t)\tilde{a}^*_{\alpha\alpha}(t) \ket{\alpha\beta}\bra{\beta\alpha} + |\tilde{a}_{\alpha\alpha}(t)|^2 \ket{\alpha\beta}\bra{\alpha\beta}
\end{split}
\end{equation}
\end{widetext}

The same dephasing map can be written in the standard Lindblad form used in neutrino-decoherence phenomenology. In the propagation variable \(L\simeq t\) (natural units), the density matrix obeys
\begin{equation}
\frac{d\rho}{dL}=-i[H,\rho]+\sum_n\left(L_n\rho L_n^\dagger-\frac{1}{2}\{L_n^\dagger L_n,\rho\}\right),
\label{eq:lindblad_standard}
\end{equation}
where the dissipative part is commonly parameterized by damping coefficients \(\Gamma_{ij}(E)\). For the single dephasing operator relevant to Eq.~\eqref{rod},
\begin{equation}
L_\phi=\sqrt{\frac{\Gamma_{ij}(E)}{2}}\,(\sigma_3\otimes I),
\label{eq:lindblad_dephasing_operator}
\end{equation}
the Lindblad dissipator gives \(d\rho_{23}/dL=-\Gamma_{ij}(E)\rho_{23}\) and leaves \(\rho_{22}\) and \(\rho_{33}\) unchanged. Hence
\begin{equation}
\rho_{23}(L)=e^{-\Gamma_{ij}(E)L}\rho_{23}(0),\qquad \gamma_{ij}(L,E)=1-e^{-\Gamma_{ij}(E)L}.
\label{eq:gamma_lindblad_relation}
\end{equation}
Thus the factor \((1-\gamma)\) appearing in Eq.~\eqref{rod} is exactly the usual Lindblad damping factor \(e^{-\Gamma_{ij}(E)L}\) that multiplies the interference terms in neutrino-oscillation probabilities~\cite{DeRomeri2023,Oliveira2016,CoelhoMann2017,Coloma2018}. For energy-dependent searches one may use \(\Gamma_{ij}(E)=\Gamma_{ij}(E_0)(E/E_0)^n\), with \(E_0=1\,\mathrm{GeV}\), as in the global analysis of Ref.~\cite{DeRomeri2023}. Physically, \(\Gamma_{ij}\) summarizes the integrated strength of the particular decohering mechanism under consideration, such as stochastic matter fluctuations, fluctuating backgrounds, wave-packet separation, or more exotic open-system effects. Consequently, the dephasing strength \(\gamma_{ij}\) is not a universal constant shared by KamLAND, MINOS, and Daya Bay; it is an experiment- and channel-dependent quantity determined by \(L\), \(E\), the relevant mass pair \((i,j)\), and the assumed energy dependence of \(\Gamma_{ij}\).

For compactness, we write \(\gamma\equiv\gamma_{ij}(L,E)\) in the formulas below.

Next, we compute the entanglement of formation, quantum discord, and local quantum uncertainty after applying the dephasing channel to the neutrino system, and we investigate the impact of decoherence on the three experiments.\\

In the following analysis, the dephasing strength is not chosen as an arbitrary large number. Instead, we use the experimentally constrained relation
\begin{equation}
\gamma_r(L)=1-\exp[-r\,\Gamma_{90}L],\qquad r\in\{0,0.25,0.5,1\},
\label{eq:gamma_constrained_benchmark}
\end{equation}
where \(r\) is a fraction of the current 90\% C.L. upper limit. As a representative energy-independent benchmark we take the Model-A bound \(\Gamma_{90}=5.1\times10^{-24}\,\mathrm{GeV}\) from Ref.~\cite{DeRomeri2023}. Using \(1\,\mathrm{km}=5.07\times10^{18}\,\mathrm{GeV}^{-1}\), this gives the following maximum allowed effective dephasing strengths at the characteristic baselines used in this work:
\begin{table}[h]
\centering
\caption{Representative effective dephasing strengths inferred from the energy-independent 90\% C.L. bound \(\Gamma_{90}=5.1\times10^{-24}\,\mathrm{GeV}\) of Ref.~\cite{DeRomeri2023}.}
\label{tab:gamma_constraints}
\begin{tabular}{lcc}
\hline
Experiment & Baseline used for estimate & \(\gamma_{90}=1-e^{-\Gamma_{90}L}\) \\
\hline
KamLAND & \(180\,\mathrm{km}\) & \(4.6\times10^{-3}\) \\
MINOS & \(735\,\mathrm{km}\) & \(1.9\times10^{-2}\) \\
Daya Bay & \(1.912\,\mathrm{km}\) & \(4.9\times10^{-5}\) \\
\hline
\end{tabular}
\end{table}
For other energy dependences or decoherence models, the corresponding bound on \(\Gamma_{ij}(E)\) from Ref.~\cite{DeRomeri2023} should be inserted into Eq.~\eqref{eq:gamma_lindblad_relation}.

\subsection{Entanglement of formation under dephasing channel}  
The entanglement of formation after the application of a dephasing channel can be determined by calculating the concurrence of the density matrix $\rho_{\alpha\beta}^D(t)$ given in Eq.~\ref{rod}. This corresponds to an X-state, characterized by non-zero elements only along the diagonal and anti-diagonal. For such states, the concurrence takes a simplified form:
	\begin{equation}
		C(\rho_{\alpha\beta}^{D}(t)) = 2 \max\big\{0, |\rho_{23}| - \sqrt{\rho_{11}\rho_{44}}\big\}.
		\label{eq:concurrence_Xstate}
	\end{equation}
	In our particular case, the relevant matrix elements are given by:
	\begin{align}
		\rho_{23} &= (1 - \gamma)\tilde{a}_{\alpha\alpha}(t)\tilde{a}^*_{\alpha\beta}(t), \label{eq:rho23} \\
		\rho_{11} &= \rho_{44} = 0, \label{eq:rho_diag}
	\end{align}
	where $\gamma \in [0,1]$ represents the dephasing parameter. This specific structure leads to a simplified expression for the concurrence:

	\begin{align}
		C(\rho_{\alpha\beta}^{D}(t)) &= 2 \max\big\{0, |(1 - \gamma)\tilde{a}_{\alpha\alpha}(t)\tilde{a}^*_{\alpha\beta}(t)|\big\} \notag \\
		&= 2(1 - \gamma)|\tilde{a}_{\alpha\alpha}(t)\tilde{a}^*_{\alpha\beta}(t)|. \label{eq:concurrence_simple}
	\end{align}

	Introducing the transition probabilities $P_{\alpha\alpha}(t) = |\tilde{a}_{\alpha\alpha}(t)|^2$ and $P_{\alpha\beta}(t) = |\tilde{a}_{\alpha\beta}(t)|^2$, we obtain the final form for the concurrence:

	\begin{equation}
		C\big(\rho_{AB}^{D}(t)\big) = 2(1 - \gamma)\sqrt{P_{\alpha\alpha}(t)P_{\alpha\beta}(t)}.
		\label{eq:concurrence_final}
	\end{equation}

	The entanglement of formation, which quantifies the quantum resources required to create the entangled state, is then expressed in terms of the concurrence via Wootters' formula~\cite{Wootters}:

	\begin{widetext}
\begin{equation}
		E_F(\rho_{\alpha\beta}^{D}(t)) = h\left( \frac{1}{2}\left( 1 + \sqrt{1 - 4(1-\gamma)^2P_{\alpha\alpha}(t)P_{\alpha\beta}(t) }\right) \right)
		\label{eq:entanglement_formation}
	\end{equation}
\end{widetext}

	where $h(x) = -x\log_2 x - (1-x)\log_2(1-x)$ is the binary entropy. This expression explicitly shows how dephasing (through $\gamma$) gradually suppresses the quantum entanglement of the system.

\begin{figure*}[t]
\centering
\begin{minipage}[b]{0.32\textwidth}
\centering
\includegraphics[width=\linewidth]{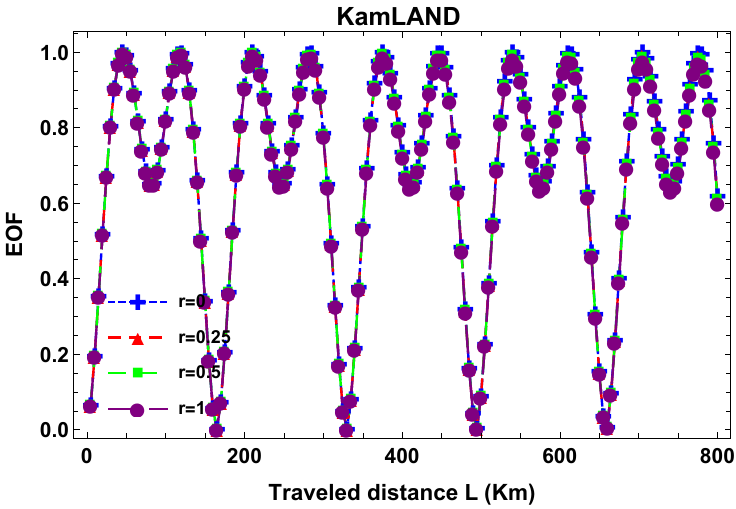}\par
(a)
\end{minipage}
\hfill
\begin{minipage}[b]{0.32\textwidth}
\centering
\includegraphics[width=\linewidth]{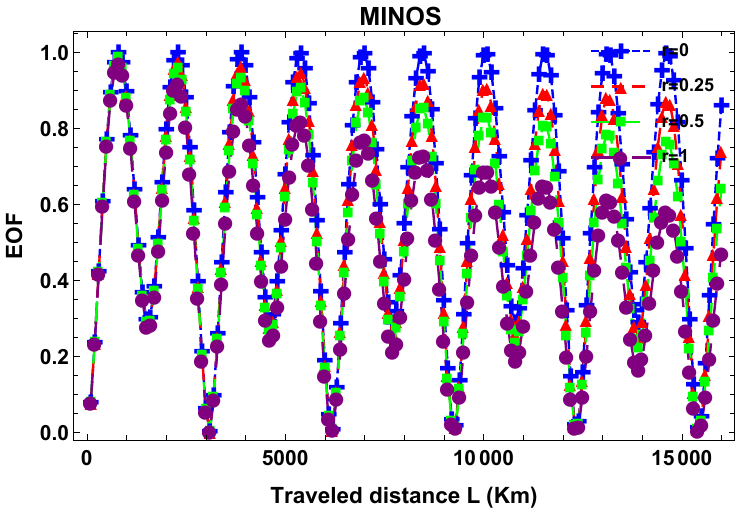}\par
(b)
\end{minipage}
\hfill
\begin{minipage}[b]{0.32\textwidth}
\centering
\includegraphics[width=\linewidth]{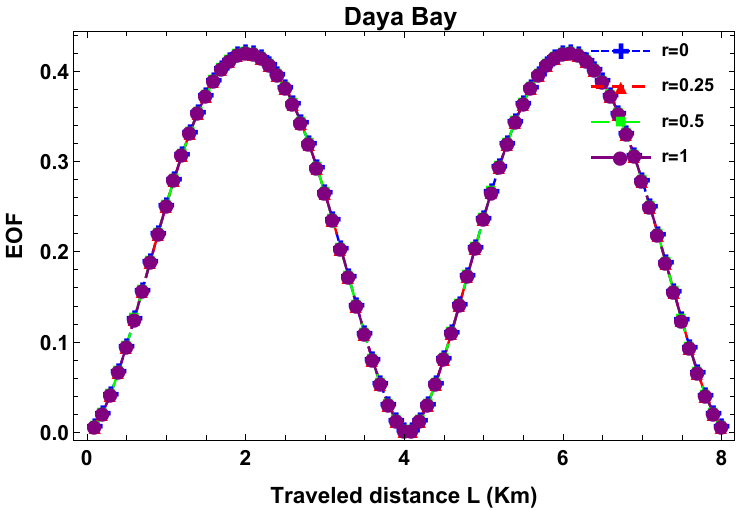}\par
(c)
\end{minipage}
\caption{Baseline dependence of the entanglement of formation under the experimentally constrained dephasing benchmark of Eq.~\eqref{eq:gamma_constrained_benchmark}, with \(r=\Gamma/\Gamma_{90}=0,0.25,0.5,1\), in (a) KamLAND, (b) MINOS, and (c) Daya Bay.}
\label{Fig5}
\end{figure*}

Figures~\ref{Fig5}(a)--(c) show the baseline dependence of the entanglement of formation after replacing the arbitrary fixed dephasing strengths by the experimentally constrained benchmark \(\gamma_r(L)=1-\exp[-r\Gamma_{90}L]\). Because the current bounds imply \(\gamma\ll1\) for the baselines considered here, the constrained curves remain close to the coherent limit, especially for the short-baseline Daya Bay case. This is physically expected: present oscillation data allow only small damping of the interference terms over these baselines. The hierarchy among KamLAND, MINOS, and Daya Bay therefore continues to be controlled mainly by the mixing angle, while the allowed Lindblad damping produces only a bounded attenuation of the EOF amplitude.

\subsection{Quantum discord under dephasing channel}  
	In the quantum discord section, we detailed the calculation of this quantifier for later use, following the same procedure to compute quantum discord under a dephasing channel.  
	The final expression for quantum discord, as previously defined, is:  
	\begin{align}  
		Q(\rho_{AB})&= \min \left[  
		H\left(\rho_{22}\right) + \lambda_{1} \log_2 \lambda_{1} + \lambda_{2} \log_2 \lambda_{2} \right.\notag\\& \left.+ 
		H\left( \frac{1 + \sqrt{(1 - 2\rho_{33})^2 + 4|\rho_{23}|^2}}{2} \right), \;  
		2|\rho_{23}|^2	\right]  
	\end{align}  
	Projecting onto the density matrix \ref{rod}, we find by analogy:  
	\begin{align*}  
		|\rho_{23}| &= (1-\gamma)\sqrt{P_{\alpha\alpha}P_{\alpha\beta}} \\  
		\rho_{22} &= P_{\alpha\beta}  \\  
		\rho_{33} &= P_{\alpha\alpha}    
	 \end{align*}  
	The eigenvalues \(\lambda_i\) become:  
	\begin{equation}  
		\lambda_{1,2} = \frac{1}{2} \left(1 \pm \sqrt{(2P_{\alpha\alpha}-1)^2 + 4(1-\gamma)^2 P_{\alpha\alpha}P_{\alpha\beta}}\right)  
	\end{equation}  
	The final quantum discord is thus:  
	\begin{widetext}
\begin{align}  
		Q(\rho_{AB}) &= \min \left[  
		\begin{aligned}  
			&H(P_{\alpha\alpha}) + \lambda_1 \log_2 \lambda_1 + \lambda_2 \log_2 \lambda_2 \\  
			&+ H \left( \frac{1 + \sqrt{(2P_{\alpha\alpha} - 1)^2 + 4(1 - \gamma)^2 P_{\alpha\alpha} P_{\alpha\beta}}}{2} \right), \\  
			&2(1 - \gamma)^2 P_{\alpha\alpha} P_{\alpha\beta}  
		\end{aligned}  
		\right]  
	\end{align}
\end{widetext}

\begin{figure*}[t]
\centering
\begin{minipage}[b]{0.32\textwidth}
\centering
\includegraphics[width=\linewidth]{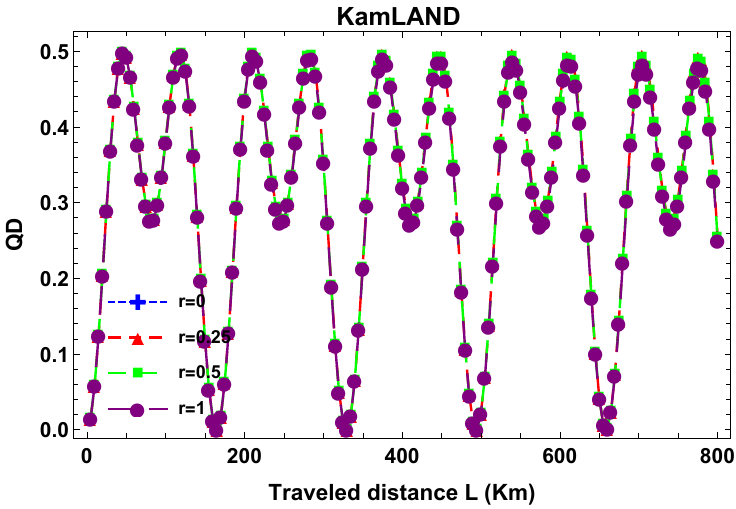}\par
(a)
\end{minipage}
\hfill
\begin{minipage}[b]{0.32\textwidth}
\centering
\includegraphics[width=\linewidth]{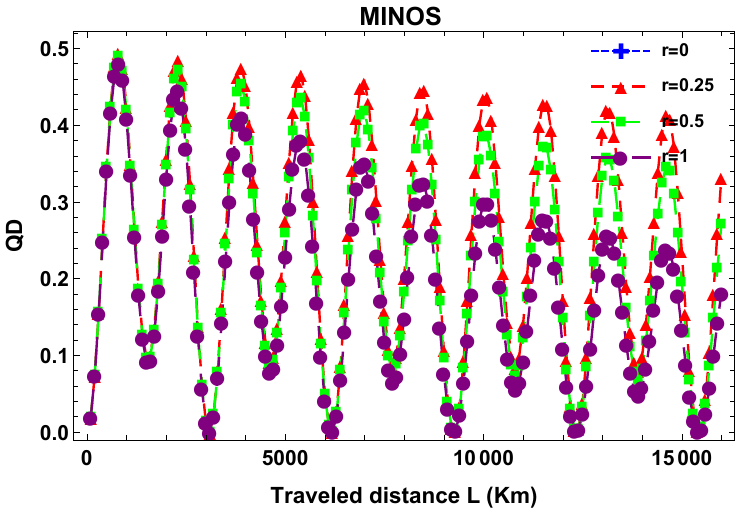}\par
(b)
\end{minipage}
\hfill
\begin{minipage}[b]{0.32\textwidth}
\centering
\includegraphics[width=\linewidth]{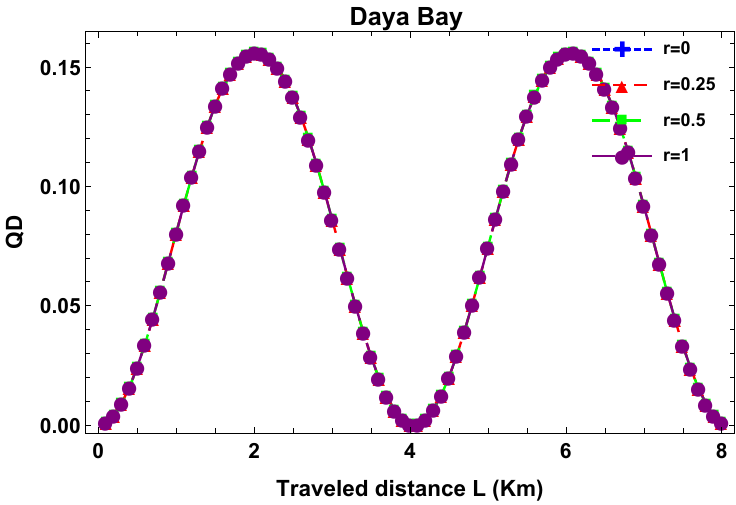}\par
(c)
\end{minipage}
\caption{Baseline dependence of quantum discord \(Q(\rho_{AB})\) under the experimentally constrained dephasing benchmark of Eq.~\eqref{eq:gamma_constrained_benchmark}, with \(r=\Gamma/\Gamma_{90}=0,0.25,0.5,1\), in (a) KamLAND, (b) MINOS, and (c) Daya Bay.}
\label{Fig6}
\end{figure*}

Figure~\ref{Fig6} shows the corresponding constrained behavior of QD. Within the experimentally allowed damping range, QD is only mildly suppressed because the interference factor \(e^{-\Gamma_{ij}L}\) remains close to unity for the representative baselines. The result is consistent with the role of QD as a broader witness of non-classical correlations: it follows the oscillation-induced redistribution of amplitudes and is reduced by Lindblad dephasing, but the presently allowed dephasing rates do not generate the large suppression that would occur for phenomenological choices such as \(\gamma=0.2\)--\(0.8\).\par

	These results align with the work of Blasone et al.~\cite{Blasone2008} and Girolami et al, who show that discord can persist even in the absence of entanglement, representing a useful resource for quantum information. Furthermore, this corroborates Bilenky's theoretical predictions~\cite{Bilenky2010} on the fundamental role of quantum correlations in neutrino oscillations. In summary, despite decoherence effects, Figure \ref{Fig6} highlights the persistence of non-classical quantum signatures, opening interesting perspectives for applications in relativistic quantum information.
\subsection{LQU under dephasing channel}
Decoherence emerges from system-apparatus-environment interactions \cite{ref45,ref46}. This section examines LQU behavior under decoherence effects. Using Eqs.~\eqref{eq:LQU_closed_form} and \eqref{eq:LQU_W_matrix}, we compute the dephased matrix \(W_{DC}\) to determine LQU:
\begin{equation}
	W_{DC} = \operatorname{diag}\left( w_{DC}^{11},\; w_{DC}^{22},\; w_{DC}^{33} \right)
\end{equation}
with matrix elements:
\begin{align}
    w_{DC}^{11} &= w_{DC}^{22} = 0,
\end{align}
\begin{widetext}
\begin{align}
	w_{DC}^{33} = &\left| \tilde{a}_{\alpha\alpha}(t) \right|^2 \left( 1 - \frac{4 (\gamma-1)^2 \left| \tilde{a}_{\alpha\beta}(t) \right|^2}{1 + 2 \sqrt{(2-\gamma)\gamma} \, \left| \tilde{a}_{\alpha\alpha}(t) \right| \left| \tilde{a}_{\alpha\beta}(t) \right|} \right)\notag\\
	&+\left| \tilde{a}_{\alpha\beta}(t) \right|^2 .
\end{align}
\end{widetext}

The LQU expression then becomes:
\begin{align}
	\mathcal{U}(\rho) &= 1 - \max\left[ 0,\; \left| \tilde{a}_{\alpha\beta}(t) \right|^2 + \left| \tilde{a}_{\alpha\alpha}(t) \right|^2 \right. \notag \\
	&\left. \times \left( 1 - \frac{4 (\gamma-1)^2 \left| \tilde{a}_{\alpha\beta}(t) \right|^2}{1 + 2 \sqrt{(2-\gamma)\gamma} \, \left| \tilde{a}_{\alpha\alpha}(t) \right| \left| \tilde{a}_{\alpha\beta}(t) \right|} \right) \right]
\end{align}
Equivalently, in terms of transition probabilities:
\begin{align}
	\mathcal{U}(\rho) &= 1 - \max\left[ 
	0,\; P_{\alpha\beta}(t) + P_{\alpha\alpha}(t) \times\right.\notag\\&\left. \left(1 - 
	\frac{4(\gamma-1)^2 P_{\alpha\beta}(t)}{1 + 2 \sqrt{(2-\gamma)\gamma\, P_{\alpha\alpha}(t) P_{\alpha\beta}(t)} }
	\right)
	\right]
	\label{eq:LQU_dephased_final}
\end{align}

	\begin{figure*}[t]
\centering
\begin{minipage}[b]{0.32\textwidth}
\centering
\includegraphics[width=\linewidth]{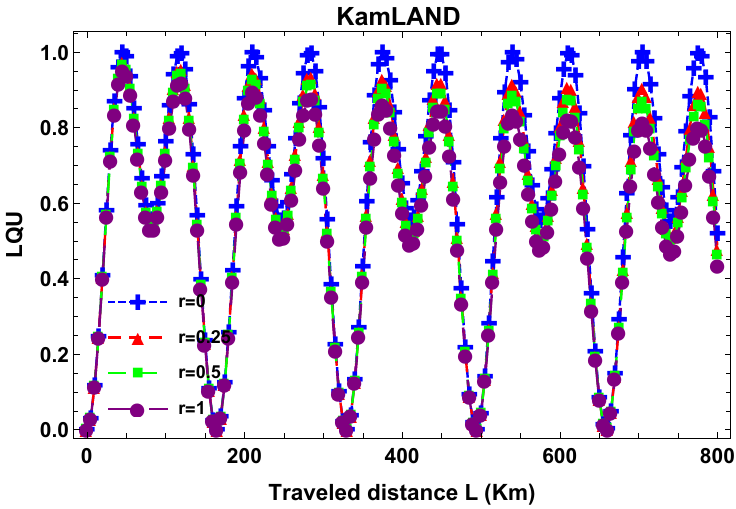}\par
(a)
\end{minipage}
\hfill
\begin{minipage}[b]{0.32\textwidth}
\centering
\includegraphics[width=\linewidth]{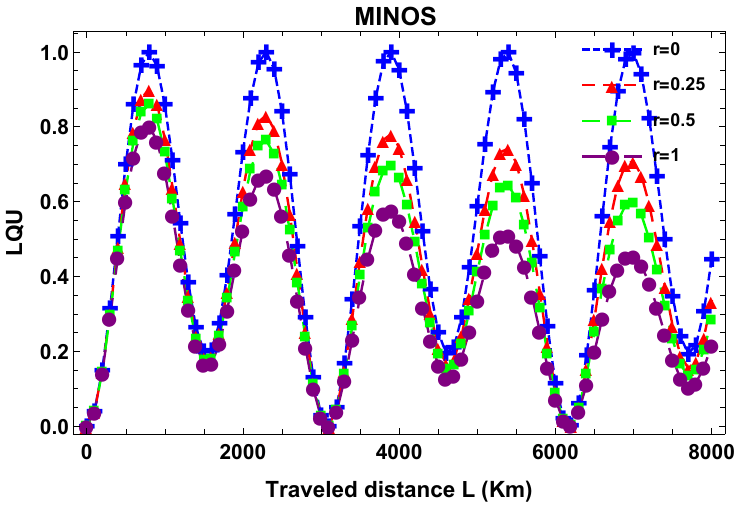}\par
(b)
\end{minipage}
\hfill
\begin{minipage}[b]{0.32\textwidth}
\centering
\includegraphics[width=\linewidth]{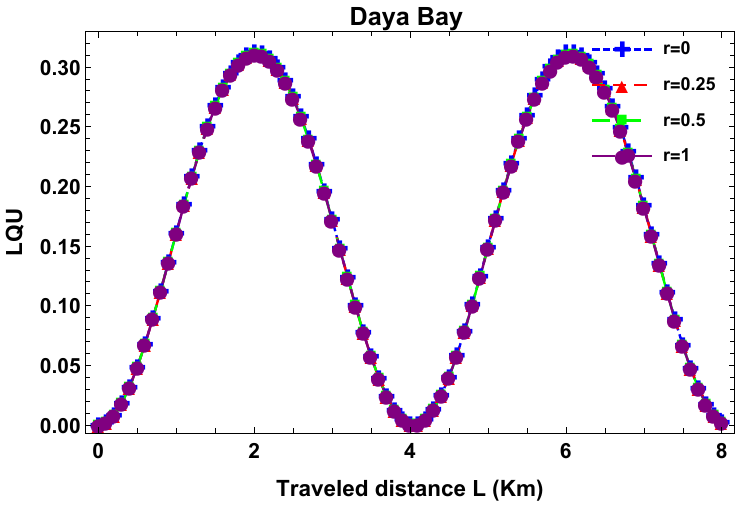}\par
(c)
\end{minipage}
\caption{Baseline dependence of LQU \(\mathcal{U}(\rho_{AB})\) under the experimentally constrained dephasing benchmark of Eq.~\eqref{eq:gamma_constrained_benchmark}, with \(r=\Gamma/\Gamma_{90}=0,0.25,0.5,1\), in (a) KamLAND, (b) MINOS, and (c) Daya Bay.}
\label{Fig7}
\end{figure*}

Figure~\ref{Fig7} presents the constrained LQU baseline dependence. As for EOF and QD, replacing fixed illustrative values of \(\gamma\) by \(\gamma_r(L)=1-e^{-r\Gamma_{90}L}\) yields a small but controlled attenuation of LQU, with the strongest visible effect in the long-baseline MINOS range. This form makes the comparison with the Lindblad literature direct, since increasing \(r\) corresponds to increasing the dissipative parameter \(\Gamma_{ij}\) up to its current 90\% C.L. upper bound.\par

Comparatively, MINOS (long baseline) shows remarkable persistence of LQU values over extended timescales, suggesting that quantum coherence effects can be maintained over long distances in certain energy regimes \cite{Blasone2014}. Conversely, Daya Bay (short baseline) exhibits faster dynamics with globally reduced amplitudes, indicating enhanced sensitivity to environmental effects in short-distance configurations.

These observations corroborate theoretical expectations for quantum correlations in open systems: the allowed damping reduces the off-diagonal coherence terms, but current bounds keep the effective damping small over the representative baselines. The marked differences between the three experiments, therefore, mainly reflect their mixing angles and \(L/E\) ranges, whereas the Lindblad benchmark quantifies the maximum coherence suppression currently allowed.

\section{Concluding Remarks}

This study presents a systematic analysis of quantum-correlation dynamics in two-flavor neutrino oscillations using three complementary quantifiers: entanglement of formation (EOF), quantum discord (QD), and local quantum uncertainty (LQU). By applying these measures to experimental parameters from KamLAND, MINOS, and Daya Bay, we have identified several key features that refine our understanding of quantum-information aspects in neutrino physics.

Our results show that neutrino oscillations naturally generate bipartite quantum correlations whose baseline dependence is governed by the interplay between flavor mixing and mass–eigenstate superposition. These correlations exhibit oscillatory behavior synchronized with flavor transition probabilities, with amplitudes set primarily by the magnitude of the mixing angle rather than by the baseline length alone. A cross-experimental comparison reveals a clear hierarchy: MINOS achieves near-maximal correlations (EOF, LQU $\approx 1$) due to near-maximal atmospheric mixing ($\theta_{23} \simeq 45^\circ$), which enables strong flavor admixture between $\nu_\mu$ and $\nu_\tau$ states. KamLAND displays intermediate correlation strengths ($\sim 0.95$), reflecting the non-maximal solar mixing angle ($\theta_{12} \approx 33.5^\circ$), which prevents complete flavor conversion even at oscillation maxima. Daya Bay exhibits the weakest correlations (EOF, LQU $\sim 0.45$), corresponding to the smallest mixing parameter ($\theta_{13} \approx 8.5^\circ$). Crucially, our analysis clarifies that the maxima of quantum correlations occur at points of maximum flavor admixture, when the neutrino state contains balanced superpositions of flavor eigenstates, rather than at oscillation maxima where a single flavor dominates. This distinction is particularly pronounced in MINOS, where maximal mixing allows nearly complete flavor conversion, causing the correlation measures to reach minima precisely when the transition probability peaks.

While the baseline length ($L$) and neutrino energy ($E$) optimize experimental sensitivity to specific mass-squared differences through the resonance condition $\Delta m^2 L / 4E \approx \pi/2$, the mixing angle remains the primary determinant of the achievable degree of quantumness. This insight corrects interpretations that attribute the correlation strength solely to geometric or kinematic factors and instead establishes that the mixing parameters are fundamental quantum-information resources encoded in the neutrino sector.

The sensitivity analysis in Sec .~\ref {subsec:sensitivity} also clarifies the metrological interpretation. For standard two-flavor oscillation parameters, the pure-state correlation measures are deterministic nonlinear functions of \(P_{\alpha\beta}\). Hence, they do not, by themselves, increase the Fisher information available from an ideal full-probability measurement. Their complementary role is instead to expose the off-diagonal coherence content of the state. In the pure-dephasing channel, \(\partial P_{\alpha\beta}/\partial\gamma=0\), whereas EOF, QD, LQU, and concurrence respond to \(\gamma\) through the suppressed coherence \(|\rho_{23}|\). This makes the correlation measures useful diagnostics of decoherence even when population probabilities are unchanged by the phase-damping map.

The investigation of decoherence effects via a dephasing channel reveals the robustness of quantum correlations under environmental perturbations. Coherence suppression is modeled through the Lindblad-connected factor \(1-\gamma_{ij}(L,E)=e^{-\Gamma_{ij}(E)L}\), with \(\Gamma_{ij}\) chosen within current experimental limits rather than as an unconstrained phenomenological constant. Within these bounds, the amplitudes of all three quantifiers are reduced in a controlled way and remain close to the coherent limit for the representative baselines considered here. This resilience shows that quantum signatures can persist in open-system regimes, supporting the view of neutrinos as macroscopic quantum systems in which non-classical features survive over kilometer-scale baselines and GeV energy scales. The persistence of quantum discord in regimes where entanglement vanishes further confirms that non-classical correlations extend beyond traditional entanglement-based frameworks.

A comparative analysis of the three correlation measures highlights their complementarity and distinct physical content. EOF and LQU exhibit closely related behavior in pure-state regimes, both peaking at maximum flavor admixture. In particular, LQU satisfies the exact relation $\mathcal{U}=\mathcal{C}^2$ for the pure two-qubit states considered here. At the same time, EOF remains a monotonic function of the concurrence rather than an identical quantity. Quantum discord, however, captures residual quantum correlations in mixed states and low-entanglement configurations, serving as a more general witness of non-classicality. This multi-measure approach is therefore essential for a comprehensive characterization of quantum resources, as single quantifiers may overlook subtle correlation structures that are particularly relevant in decoherence-affected or weakly mixing regimes.

Our findings support the view that neutrino oscillations provide a useful theoretical laboratory for relativistic quantum-information concepts. Within the limitations of an effective two-flavor, parameter-based treatment, the present analysis should be regarded as a controlled proof of principle rather than as a direct experimental constraint study. The framework may nevertheless be useful for motivating future work on quantum Fisher information in oscillation physics, on open-system tests of coherence loss, and on phenomenological searches for non-standard damping effects.

Future investigations should extend this framework to three-flavor oscillations, including CP-violating effects, explore matter-induced modifications of quantum correlations in solar and atmospheric neutrino propagation, and investigate the role of quantum resources in distinguishing between standard oscillations and exotic scenarios such as non-standard interactions or quantum-gravity effects. The demonstrated resilience of quantum correlations under decoherence also motivates experimental searches for deviations from standard coherence predictions, potentially constraining new physics beyond the Standard Model through precision correlation tomography in next-generation neutrino experiments.

\appendix
\section{Derivation of the local quantum uncertainty for the two-flavor state}
\label{app:lqu_derivation}

This appendix gives the algebraic steps leading to Eq.~\eqref{eq:LQU_final}. The density matrix \(\rho^{\alpha\beta}(t)=|\nu_\alpha(t)\rangle\langle\nu_\alpha(t)|\) has support only on the subspace spanned by \(|\beta\alpha\rangle\) and \(|\alpha\beta\rangle\). Since the state is pure, \(\sqrt{\rho^{\alpha\beta}}=\rho^{\alpha\beta}\). Substituting this projector into the two-qubit LQU matrix
\begin{equation}
W_{ij}=\operatorname{Tr}\left[\sqrt{\rho^{\alpha\beta}}\,(\sigma_i\otimes I_2)\sqrt{\rho^{\alpha\beta}}\,(\sigma_j\otimes I_2)\right]
\end{equation}
leaves only one non-zero eigenvalue,
\begin{equation}
\lambda_3=\left(|\tilde a_{\alpha\alpha}(t)|^2-|\tilde a_{\alpha\beta}(t)|^2\right)^2,
\qquad \lambda_1=\lambda_2=0.
\end{equation}
Using \(|\tilde a_{\alpha\alpha}(t)|^2+|\tilde a_{\alpha\beta}(t)|^2=1\), the LQU becomes
\begin{align}
\mathcal{U}(\rho^{\alpha\beta}(t))
&=1-\left(|\tilde a_{\alpha\alpha}(t)|^2-|\tilde a_{\alpha\beta}(t)|^2\right)^2\notag\\
&=4|\tilde a_{\alpha\alpha}(t)|^2|\tilde a_{\alpha\beta}(t)|^2\notag\\
&=4P_{\alpha\alpha}(t)P_{\alpha\beta}(t).
\end{align}
This is exactly the squared concurrence of the pure effective two-qubit oscillation state.

\end{document}